\documentclass[12pt,letterpaper]{article}
\pdfoutput=1
\usepackage{graphicx}
\usepackage{caption}
\usepackage{subcaption}

\usepackage{amsfonts,amssymb,amsmath}

\newcommand{\be}{\begin{equation}}
\newcommand{\ee}{\end{equation}}
\newcommand{\ben}{\begin{displaymath}}
\newcommand{\een}{\end{displaymath}}
\newcommand{\bal}{\begin{align}}
\newcommand{\eal}{\end{align}}
\newcommand{\bean}{\begin{eqnarray*}}
\newcommand{\eean}{\end{eqnarray*}}

\newcommand{\ads}[1]{\mbox{${AdS}_{#1}$}}
\newcommand{\adss}[2]{\mbox{$AdS_{#1}\times {S}^{#2}$}}

\newcommand{\eg}{{\it e.g.}}
\newcommand{\ie}{{\it i.e.}}

\newcommand{\tr}{\mbox{Tr}}

\newcommand{\commentout}[1]{}

\newcommand{\X}{\ensuremath{\mathbb{X}}}

\newcommand{\pds}{\ensuremath{\partial_\sigma}}
\newcommand{\pdt}{\ensuremath{\partial_\tau}}

\newcommand{\beq}{\begin{equation}}
\newcommand{\eeq}{\end{equation}}
\newcommand{\beqa}{\begin{eqnarray}}
\newcommand{\eeqa}{\end{eqnarray}}
\newcommand{\beqar}{\begin{eqnarray*}}
\newcommand{\eeqar}{\end{eqnarray*}}
\newcommand{\cN}{{\cal N}}

\newcommand{\cO}{{\cal O}}

\newcommand{\non}{\nonumber}

\newcommand{\cA}{{\cal A}}
\newcommand{\cB}{{\cal B}}

\newcommand{\half}{\ensuremath{\frac{1}{2}}}

\newcommand{\bz}{{\ensuremath{\bar{z}}}}

\newcommand{\N}[1]{\ensuremath{\cN=#1}}

\newcommand{\bpartial}{\ensuremath{\bar{\partial}}}

\newcommand{\cU}{\ensuremath{\mathcal{U}}}

\newcommand{\lv}{\ensuremath{\int_1^4}}
\newcommand{\lvt}{\ensuremath{\int_1^5}}
\newcommand{\thetah}{\ensuremath{\hat{\theta}}}
\newcommand{\hR}{\ensuremath{\hat{R}}}
\newcommand{\hRi}{\ensuremath{\hat{R}^{-1}}}
\newcommand{\lvo}{\int_1^4}

\begin{document}

\title{\LARGE \bf Euclidean Wilson loops and Minimal Area Surfaces in Minkowski \ads{3}}

\author{Andrew Irrgang$^{1,2}$, Martin Kruczenski$^2 $ \thanks{E-mail: \texttt{airrgang1@ivytech.edu, markru@purdue.edu,}}\\
        $^1$ Ivy Tech Community College, Lafayette, IN \\
       $^2$ Dep. of Physics and Astronomy, \\ Purdue University, W. Lafayette, IN   }

\maketitle

\begin{abstract}
 The AdS/CFT correspondence relates Wilson loops in \N{4} SYM theory to minimal area surfaces in \adss{5}{5} space.
If the Wilson loop is Euclidean and confined to a plane $(t,x)$ then the dual surface is Euclidean and 
lives in Minkowski $\ads{3}\subset \ads{5}$. In this paper we study such minimal area surfaces generalizing previous results
obtained in the Euclidean case. Since the surfaces we consider have the topology of a disk, the holonomy of the flat current vanishes which is equivalent to the condition that a certain boundary Schr\"odinger equation has all its solutions anti-periodic. If the potential for that Schr\"odinger equation is found then reconstructing the surface and finding 
the area become simpler. In particular we write a formula for the Area in terms of the Schwarzian derivative of the 
contour. Finally an infinite parameter family of analytical solutions using Riemann Theta functions is described. 
In this case, both the area and the shape of the surface are given analytically and used to check the previous results.  
\end{abstract}

\clearpage
\newpage



\section{Introduction}

 The Wilson loop operator is one of the most fundamental operators of a gauge theory.  Its expectation value distinguishes a confining theory from one that is non-confining, is used to compute the quark/anti-quark potential, and determines the expectation value of gauge invariant operators as well as their correlation functions in various limits.  Analytical methods to compute Wilson loops in the large N-limit \cite{largeN} and for case of strong 't Hooft coupling proceeds by utilizing the AdS/CFT correspondence \cite{malda} whenever applicable.  To leading order in strong coupling, the Wilson loop is computed by finding a minimal area surface in a higher dimensional space \cite{MRY}.  For the standard case of \N{4} SYM, considered in this paper, the minimal area surfaces live in \adss{5}{5}.  This case is of particular interest because the dual string theory is described by an integrable model \cite{BPR}. Consequently, the relationship between Wilson loops and minimal area surfaces has motivated much work in the area \cite{WLref}. The most studied one is the circular Wilson loop \cite{cWL} including small perturbations around it \cite{SY,CHMS,Cagnazzo}. Also, a particularly important role has been played by Wilson loops with light-like cusps \cite{cusp} due to their relation with scattering amplitudes \cite{scatampl,AMSV}. More recently new results for Wilson loops of more general shape have started to appear \cite{KZ,Janik,IKZ,IK,Dekel}, which includes solutions using Riemann theta functions. Such solutions were obtained using the methods of \cite{BB,BBook} and similar techniques that had been previously used to find closed string solutions \cite{ClosedStrings}.
It is also important to recall that in the large-N limit the Wilson loop in the gauge theory obeys the loop equation \cite{leq} that can also be studied within AdS/CFT \cite{PR}. 

 In this paper, further insight into the properties of the Wilson loop operator is gained through study of the minimal area surfaces in \ads{5}.  Such surfaces are obtained utilizing the simple but powerful Pohlmeyer \cite{Pohlmeyer}\footnote{See \eg\ \cite{Hoare:2012nx} for a more recent description of the method.} reduction. Beginning from a Euclidean world-sheet living in $\ads{3}\subset \ads{5}$ the surface is parameterized by the complex coordinate $z$ using conformal gauge.  The world-sheet metric then reads
\beq
 ds^2 = 4 e^{2\alpha} dz\, d\bar{z}.
 \label{a1}
\eeq
Here $\alpha(z,\bar{z})$ is a real function on a region of the complex plane that can be taken as the unit disk by a conformal transformation. Further, an important observation is that $\alpha(z,\bar{z})$, the conformal factor of the world-sheet metric, obeys a non-linear equation similar to the sinh-Gordon equation,
\beq
 \partial \bar{\partial} \alpha = e^{2\alpha} - f(z) \bar{f}({\bar{z}})\, e^{-2\alpha},
 \label{a2}
\eeq
where $f(z)$ is an unknown holomorphic function. Such an equation is solvable independent of the other variables and yields that finding a minimal area surface means solving a set of linear differential equations once a solution is obtained for $\alpha(z,\bar{z})$. Further, the linear equations are deformable by a complex parameter $\lambda$ called the spectral parameter. When $|\lambda|=1$ a one-parameter family of minimal area surfaces is obtained which all have the same area. Such deformations are called $\lambda$ deformations\footnote{This name was introduced in \cite{Dekel}.} and lead to an infinite number of conserved quantities given by the holonomy of certain associated currents around a non-trivial loop on the world-sheet.  
 
 One can use the Pohlmeyer reduction in two different ways. The first one is to find new minimal area surfaces. Thus, an arbitrary function $f(z)$ is chosen and then the solution for the conformal factor is found and used to construct a surface.  The Wilson loop where the surface ends is then determined as part of the procedure. For example, an infinite parameter family of solutions were found in \cite{IKZ,KZ,IK} for the case where $f(z)$ does not vanish anywhere on the surface.  These solutions are analytic and can be written in terms of Riemann theta functions. The second way to use this method, is to try to find a minimal area surface ending in any arbitrary given curve.  The specified curve is used to compute the boundary conditions for $f(z)$ and $\alpha$ from which those functions, and the corresponding surface, can be reconstructed. For the Euclidean case, this was discussed in \cite{WLMA} where it was found that the Schwarzian derivative of the contour with respect to the conformal angle\footnote{If we write $z=r e^{i\theta}$ then $\theta$ is defined as the conformal angle parameterizing the world-sheet boundary at $r=1$.} determines all the boundary conditions necessary to reconstruct the surface. However, finding the correct parameterization of the contour in terms of the conformal angle requires solving a non-trivial problem involving reconstructing a potential depending on the spectral parameter such that all its solutions are antiperiodic \cite{WLMA}. 
 
 At this moment it is not clear how to solve such a problem but in a recent important paper by Dekel \cite{Dekel} it was shown that such problem is solvable by studying perturbations around the circle. Although such a perturbative approach had been considered before \cite{SY}, in \cite{Dekel} new methods  extend the expansion to much higher orders than before providing a useful tool for solving the problem.. 

 Another, related approach is to extend the results associated with light-like cusps \cite{AMSV} by considering the limit where the number of cusps goes to infinity in such a way that a smooth curve is reproduced. This approach is used to great effect in a recent paper by J. Toledo \cite{Toledo} where he managed to obtain a Y-system type of equation for the cross ratios associated with a given curve. The Y-system uses as an input a curve in the world-sheet describing the world-sheet boundary in the world-sheet coordinates where $f(z)=1$. In the language of the Pohlmeyer reduction this is equivalent to giving $f(z)$ in the 
 coordinates where the world-sheet is the unit disk. Instead of using the more difficult approach of solving for $\alpha$ and then computing the area, Toledo showed that, from the solution to the Y-system of equations, the shape of the Wilson loop and the area of the associated surface follow. As mentioned before, this approach was derived in a roundabout way and a direct derivation that connects it with the methods discussed here and in \cite{WLMA} would make the discussion more complete. 
 
 This paper is organized as follows: In the next section the Pohlmeyer reduction is implemented for this case. In section 3 we make a construction analogous to \cite{WLMA}. We argue that the Schwarzian derivative of the contour with respect to the conformal angle gives the necessary boundary conditions for the functions $f(z)$ and $\alpha(z,\bar{z})$. Again, the conformal angle is found in principle by requiring that all conserved charges vanish or, equivalently, by finding a potential in a boundary Schr\"odinger equation such that the solution to that equation is anti-periodic for all values of the spectral parameter. Finally we give a formula for the area that is valid when $f(z)$ has no zeros in the unit disk\footnote{This condition also applies to the formula given in \cite{WLMA} although it was not made explicit there.}  In section 4 we present an infinite parameter family of solutions in terms of Riemann theta functions
 following similar methods of \cite{IKZ,KZ,IK}. These new solutions correspond to the case where $f(z)$ has no zeros in the unit disk and can be used to test the results of the previous sections. They should also be useful to test the results of \cite{Toledo} but we leave that for future work. Finally, in the last section, we give our conclusions. In an appendix we collect several useful formulas for theta functions and perform the computation of the Schwarzian derivative of the contour in terms those functions.

\section{Integrability and Pohlmeyer reduction}

 Surfaces of minimal area are found by implementing the well-known Pohlmeyer Reduction \cite{Pohlmeyer} which is based on the integrability of the string Sigma Model.  The utility of the method is due to its simplification of the problem; namely, it reduces solving the non-linear string equations of motion (including the conformal constraints) to solving a single Sinh-Gordon equation plus a set of linear differential equations.

 This work builds upon previous results found in \cite{IK} by again considering general open string solutions in \ads{3} when embedded in Minkowski space but now for the case of a world--sheet with Euclidean signature.  With the Minkowski embedding in mind, the manifold \ads{3} is defined as a subspace of $R^{2,2}$ subject to a constraint on the coordinates $X^\mu$ ($\mu = -1,0,1,2$),
\beq
X_\mu X^\mu =  -X_0^2 - X_{-1}^2 +X_1^2 + X_2^2 = -1.
 \label{Ads}
\eeq
For later convenience, the relationships between the embedding coordinates and global coordinates $(t,\phi,\rho))$ and Poincare coordinates are now defined through the expressions (\ref{a3}) and (\ref{a4}) respectively.
\begin{align}
	X_{-1}+iX_0 = \cosh\rho \, e^{it}, \ \ \ \ X_1+i X_2 = \sinh\rho\, e^{i\phi}
		\label{a3} \\
	Z=\frac{1}{X_{-1}-X_2}, \ \ \ X = \frac{X_1}{X_{-1}-X_2}, \ \ \ 
	T=\frac{X_0}{X_{-1}-X_2}
		\label{a4}
\end{align}

Further, the world--sheet is parameterized by the conformal coordinates $(\sigma,\tau)$ or equivalently by the complex combinations $z=\sigma+i\tau$ and $\bar{z}=\sigma-i\tau$ which are more useful for this work.  For this choice, the  world--sheet metric has the form
\beq
	ds^2 = \frac{\Lambda(z,\bar{z})}{2}\, dz\,d\bar{z}.
	\label{a7}
\eeq 
Working in conformal gauge, the action for the string Sigma Model is given by
\begin{equation}
	S = \frac{T}{2} \,  \int d\tau \, d\sigma \, (\pds{X}^\mu \pds{X}_\mu + \pdt{X}^\mu \pdt{X}_\mu  + \Lambda(X^\mu X_\mu + 1))
	\label{a5}
\end{equation}
where the Lagrange multiplier $\Lambda$ enforces the embedding constraint.  Consequently, following from the action and the gauge choice, the equations (\ref{eom1}), (\ref{a6a}), and (\ref{a6b}) determine a surface of minimal area describing the string.
\begin{align}
	\partial_\sigma^2{X}^\mu + \partial_\tau^2{X}^\mu = \Lambda X^\mu,
		\label{eom1} \\ 
	\partial_\tau{X}^\mu\partial_\sigma{X}_\mu = 0
		\label{a6a} \\ 
	\pdt{X}^\mu \pdt{X}_\mu = \pds{X}^\mu\pds{X}_\mu
		\label{a6b}
\end{align}

Proceeding, the equations (\ref{eom1})-(\ref{a6b}) are reduced to a single Sinh-Gordon equation.  The procedure utilized here begins by forming a $2\times 2$ real matrix $\X$ using particular combinations of the embedding coordinates,
\beq
	\X = \left(\begin{array}{cc}X_{-1}+X_2&X_1+X_0\\ X_1-X_0&X_{-1}-X_2\end{array}\right)\, .
	\label{a19}
\eeq
A result of choosing these combinations is that the embedding constraint requires that $\det\X = 1$ namely $\X\in SL(2,\mathbb{R})$.  Further, any such matrix can be written as the product of any other two $SL(2,\mathbb{R})$ group elements $A_a (a=1,2)$ .  Convenient for the current work, this product is defined by the expression
\begin{equation}
	\mathbb{X} = A_1 A_2^{-1}
		\label{a8}.
\end{equation}
A useful consequence of this choice, used later, is the introduction of a redundancy in the description of $\X$ which implies an invariance under a world--sheet gauge transformation,
\beq
	A_a \rightarrow A_a\, \cU(z,\bz).
		\label{a20} 
\eeq
In addition, these two group elements are used to define two one-forms,
\begin{equation}
	J_a = A_a^{-1} d A_a,  \ \ \ a=1,2\, ,
		\label{a9}
\end{equation}
which satisfy the relationships (\ref{a10a}) and (\ref{a10b}) where no summation on $a$ is implied.
\begin{align}
	\tr J_a =0 
		\label{a10a} \\
	dJ_a +J_a\wedge J_a =0 ,
		\label{a10b}
\end{align}
For reference, the conventions used for differential forms in coordinates $z$ and $\bz$ are given by (\ref{a11})-(\ref{c1}).
\begin{align}
	a &= a_z dz + a_{\bz} d\bz, \label{a11} \\ 
	da &= (\partial a_{\bz}- \bar{\partial} a_z) \, dz\wedge d\bar{z}, \label{a12} \\
	a\wedge b &= (a_z b_{\bz} - a_{\bz} b_z ) dz\wedge d\bar{z}, \label{a13}
\end{align}
\begin{equation}
  (*a)_z = -i a_z, \ (*a)_{\bz} = i a_{\bz}, \ \ *a\wedge b = -a\wedge *b, \ \ \ **a=-a.
  \label{c1}
\end{equation}

The system of equations (\ref{eom1})-(\ref{a6b}) describing the string are expressible in terms of the matrix $\X$ as shown in (\ref{c2})-(\ref{c2b}).
\begin{align}
	d\!*\!d\mathbb{X} &= \frac{i\Lambda}{2}\, \mathbb{X}\, dz\wedge d\bar{z},
		\label{c2} \\
	\det(\bar{\partial} \X) &= 0,
		\label{c2a} \\
	\det(\partial \X) &= 0.
		\label{c2b}
\end{align}
However, more relevant now are their expressions in terms of the currents $J_a$.  For the equation of motion (\ref{c2}), substitution of the currents yields 
\begin{equation}
	J_1 \wedge *J_1 + *J_1 \wedge J_2 - J_1\wedge *J_2 - *J_2\wedge J_2 + d*J_1 - d*J_2 = \frac{i\Lambda}{2} dz\wedge d\bar{z}
		\label{c3}
\end{equation}
which is simplified by the fact that the currents $J_a$ are traceless, (\ref{a10a}),
\begin{equation}
	d*(J_1-J_2) + *J_1\wedge J_2 + J_2\wedge *J_1 = 0 .
		\label{c5}
\end{equation}
In terms of the currents, the system of equations to be solved are the equations of motion and conformal constraints (\ref{c8})-(\ref{c10}) as well as the defining equations for the currents (\ref{c6}) and (\ref{c7}).
\begin{align}
	dJ_1 + J_1\wedge J_1 &= 0    \label{c6}\\
	dJ_2 + J_2\wedge J_2 &= 0     \label{c7}\\
	d*(J_1-J_2) + *J_1\wedge J_2 + J_2\wedge *J_1 &= 0  \label{c8} \\
	\det (J_{1\bz} - J_{2\bz}) &= 0 \label{c9} \\
	\det (J_{1z} - J_{2z}) &= 0 \label{c10}
\end{align}

Inspection of these equations reveals a more convenient description by defining two new currents.
\begin{align}
 \cA = \half (J_1-J_2) \label{b1a} \\
	\cB = \half(J_1+J_2) \label{b1b} 
\end{align}
Summarizing, the system of equations to solve are the following.
\begin{align}
	d\cA + \cA \wedge \cB + \cB\wedge \cA &= 0           \label{a14}\\
	d(*\cA) + (*\cA) \wedge \cB + \cB\wedge (*\cA) &= 0  \label{a15}\\
  d\cB + \cB \wedge \cB + \cA \wedge \cA &= 0          \label{a16}\\
  \det(\cA_z) = \det(\cA_\bz) &= 0                     \label{a17}\\
  \tr \cA &= 0 																				 \label{a17a}\\		
	\tr \cB &= 0                                  			 \label{a17b}
\end{align}
 
While seemingly more complicated, everything is now in place to complete the reduction and solve the problem.  A flat current $a$ is defined as a linear combination of the currents $\cA$ and $\cB$ which is also traceless.
\begin{align}
	a = \alpha \cA + \beta *\!\cA + \gamma \cB   \label{c10}\\
	da+a\wedge a &= 0   \label{c11} \\
	\tr(a) &=0 .  \label{c12}
\end{align}

The importance of the current $a$ is the realization that a one parameter family of non-trivial solutions exists given by $\alpha^2+\beta^2=1$ and $\gamma=1$.
This family is parameterized in terms of the spectral parameter $\lambda$ for which $\alpha+i\beta=i\lambda$ and $\alpha-i\beta=\frac{1}{i\lambda}$. Using these facts, the flat current is written as follows.
\begin{align}
	a &= \frac{i}{2}  \left(\lambda -\frac{1}{\lambda}\right) \cA + \half  \left(\lambda +\frac{1}{\lambda}\right) (*\cA) + \cB  \label{c13}\\
	  &= i \lambda \cA_\bz \, d\bz + \frac{1}{i\lambda} \cA_z dz + \cB.
	\label{b3} 
\end{align}
An additional restriction must be imposed since $\cA$ and $\cB$ are real whereas $\lambda$ is generically complex which means the flat current $a$ also satisfies the following reality condition.
\beq
	\overline{a(\lambda)} = a(\frac{1}{\bar{\lambda}})
	\label{b4} 
\eeq
Note that the original currents $J_a$ can be recovered using the newly defined current $a$:  $J_1= a(1)$ $J_2=a(-1)$.

To deterimine $a$, first expand the current $\cA$ in terms of the Pauli matrices, $\sigma_{a=1,2,3}$, and generically complex coefficients $n_i$ using the notation $\bar{n}_i=n_i^*$.
\begin{align}
	\cA_{\bz} &= n_1 \sigma_1 +n_2 i\sigma_2 + n_3 \sigma_3  ,  \label{c14}\\
	\cA_z &= \bar{n}_1 \sigma_1 +\bar{n}_2 i\sigma_2 + \bar{n}_3 \sigma_3 ,
	\label{b5} 
\end{align}
In this way, the conditions $\det \cA_z=0$ and $\det \cA_{\bz}=0$ are reinterpreted as a condition that the coefficients are the components of a light-like vector defined by the metric $\mbox{diag}(-,+,-)$:
\beq
	n_2^2 - n_1^2 -n_3^2 =0 . \label{b6}.
\eeq
For the coefficients written generically as $n_i = n_{i,R} + i n_{i,I}$, the above requirement produces two conditions on the real and imaginary parts:
\beq
	n_{R}^2 = n_{I}^2 \qquad	n_R.n_I = 0.
	\label{ba6} 
\eeq
Since the real and imaginary parts of the coefficient vector have the same signature and are orthogonal they must be proportional to each other and are both either space-like or light-like.

Now the gauge symmetry discussed earlier, (\ref{a20}), is re-expressed in terms of $\cA$ as 
\begin{equation}
	\cA_a \rightarrow \cU(z,\bz)^{-1} \cA_a\, \cU(z,\bz)
\end{equation}
which amounts to an $SL(2,\mathbb{R})=SO(2,1)$ rotation of the vectors $n_R$ and $n_I$. Assuming that $n^2_R\neq 0$, such a transformation always allows these vectors to be put into the following forms.
\begin{align}
	n_R = \half e^{\alpha} (0,0,1) \label{b7a} \\
	n_I = \half e^{\alpha} (1,0,0) \label{b7b}
\end{align}
In the above expressions, $\alpha(z,\bz)$ is a real function. Thus. the flat current is
\begin{align}
	a_z &= \frac{1}{i\lambda}\cA_z + \cB_z   = \frac{1}{2i\lambda} e^\alpha (i\sigma_1+\sigma_3) + \cB_z \label{b8a}\\
	a_\bz &= i\lambda \cA_\bz + \cB_\bz      = \frac{i\lambda}{2} e^{\alpha} (-i\sigma_1+\sigma_3) + \cB_\bz
	\label{b8} 
\end{align}
and the flatness condition of the current determines the components of $\cB$: 
\begin{align}
	\cB_z     &=  \half \partial    \, \alpha\, \sigma_2 + \half f(z)   e^{-\alpha}\, (\sigma_1+i \sigma_3) , \label{b9a}\\
    \cB_{\bz} &= -\half \bar{\partial}\, \alpha\, \sigma_2  + \half \bar{f}(\bz) e^{-\alpha}\, (\sigma_1+i \sigma_3)  ,
	\label{b9}
\end{align}
Here, $f(z)$ is an arbitrary holomorphic function.  In addition, $\alpha$ satisfies
\beq
	 \partial \bar{\partial} \alpha  = e^{2\alpha} - f \bar{f}  e^{-2\alpha} .
	\label{b10}
\eeq
At this point, conventions chosen for this work make it convenient to rotate the flat connection with the SU(2) matrix
\beq
 \tilde{a} = \hR a \hRi  ,  \ \ \
 \hR = \frac{1}{\sqrt{2}} (1+i\sigma_1), \ \ \hRi = \frac{1}{\sqrt{2}} (1-i\sigma_1), \ \ \ \hR^2 = i\sigma_1  \label{c16}
\eeq
 to put it in a simpler form:
\beq 
	\tilde{a}_z = \left(\begin{array}{cc}-\half\partial\alpha & f e^{-\alpha}\\ & \\ \frac{1}{\lambda}e^{\alpha} & \half \partial\alpha\end{array}\right), \ \ \ 
    \tilde{a}_{\bz} = \left(\begin{array}{cc}\half\bar{\partial}\alpha & \lambda e^{\alpha}\\  & \\ \bar{f} e^{-\alpha} & -\half \bar{\partial}\alpha\end{array}\right), 
	\label{b11}
\eeq
 The new flat current satisfies the reality condition
\beq
 \overline{\tilde{a}(\lambda)} = \sigma_1 \tilde{a}(\frac{1}{\bar{\lambda}}) \sigma_1
 \label{c17}
\eeq
 Since $\tilde{a}$ is flat, we can solve the linear problem
\beq
	d \Psi(\lambda;z,\bz) = \Psi(\lambda;z,\bz) \tilde{a} .
	\label{b12}
\eeq
 We can choose $\Psi(\lambda;z,\bz)$ to satisfy the reality condition
\beq
 \overline{\Psi(\lambda;z,\bz)} = i \Psi(\frac{1}{\bar{\lambda}};z,\bz) \sigma_1
 \label{c18}
\eeq
 where the factor of $i$ is chosen for convenience. With that choice, however, and 
 since $J_1= a(1)$, $J_2=a(-1)$, we can take $A_1=\Psi(1)\hR$, $A_2=\Psi(-1)\hR$ since $A_{1,2}$ turn out to be real. Thus, the solution to the non-linear problem reads
\beq
	\X = \Psi(1) \Psi(-1)^{-1} .
	\label{b13}
\eeq
Therefore, the strategy is to solve the equation for $\alpha$, replace it in the flat current, solve the linear problem, and reconstruct the solution $\X$. Actually, this procedure gives a one-parameter family of real solutions that 
can be written as
\beq
	\X (\lambda) = \Psi(\lambda) \Psi(-\lambda)^{-1} .
	\label{b14}
\eeq
for
 \beq
 |\lambda|=1
 \label{c18b}
 \eeq
 The reason is that eqs.(\ref{a14})-(\ref{b2}) are invariant under $\cA_z\rightarrow (1/\lambda) \cA_z$,  $\cA_\bz\rightarrow (1/\bar{\lambda}) \cA_\bz$ whenever $|\lambda|=1$. 
These surfaces end in different boundary contours but they all have the same regularized area that, for any value of $\lambda$, is given by \cite{scatampl}:
\beq
 A_f = -2\pi + 4 \int_D f\bar{f}\, e^{-2\alpha} d\sigma d\tau
 \label{c21}
\eeq
where the integral is over the domain $D$ of the solution.

\section{Schwarzian derivative and the condition of vanishing charges}

In \cite{WLMA} a method of approaching the problem using the condition of vanishing charges was described. In particular the area was computed in terms of the Schwarzian derivative of the contour. Those results were
derived for Euclidean \ads{3}, in this section we rewrite them for Minkowski \ads{3} to get some further insight into the surfaces. Later we are going to provide concrete solutions in term of theta functions. 

 Following \cite{WLMA}, in this section we take the world-sheet to be the unit disk in the complex plane $z$. The boundary of the disk maps to the contour in the boundary of \ads{3} and the interior of the disk maps to the surface of minimal area
 that we seek. Near the boundary the induced metric diverges implying that $\alpha\rightarrow\infty$. Introducing a coordinate
\beq
 \xi = 1-r^2 
  \label{d1}
\eeq
 we find that eq. (\ref{b10}) implies the behavior
\beq
 \alpha = -\ln \xi + \beta_2(\theta) (1+\xi)\xi^2 + \beta_4(\theta) \xi^4 + \cO(\xi^5), \ \ \ (\xi\rightarrow 0)
  \label{d2}
\eeq
 From here we can compute the leading behavior of the flat current as we approach the boundary. It is best written in terms of 
\begin{align}
 \tilde{a}_\xi &\simeq_{\xi\rightarrow 0} -\frac{\lambda}{2\xi} e^{-i\theta} \sigma_+ - \frac{1}{2\lambda\xi} e^{i\theta} \sigma_-  +\cO(\xi) \label{d3} \\
 \tilde{a}_\theta &\simeq_{\xi\rightarrow 0} -\frac{i}{\xi} \sigma_3 -\frac{i\lambda}{\xi} e^{-i\theta} \sigma_+ + \frac{i}{\lambda \xi} e^{i\theta} \sigma_- +\cO(1) \label{d4}
\end{align}
defined such that 
\beq
 \partial_\xi \Psi = \Psi \tilde{a}_\xi,  \ \ \ \   \partial_\theta \Psi = \Psi \tilde{a}_\theta
 \label{d5}
\eeq
 Defining 
\beq
 \Psi = \left(\begin{array}{cc}\psi_1 & \psi_2 \\ \tilde{\psi}_1 & \tilde{\psi}_2 \end{array}\right) 
 \label{d6}
\eeq
 where, from (\ref{b12}) and (\ref{b11}), $\psi_{1,2}$ satisfy the equations 
\begin{align}
 \partial \psi_1 &= - \half \partial\alpha \psi_1  + \frac{1}{\lambda} e^{\alpha} \psi_2   \label{d7}\\
 \partial \psi_2 &= \half \partial\alpha \psi_2  + f(z) e^{-\alpha} \psi_1   \label{d8}\\
 \bar{\partial} \psi_1 &= \half \bar{\partial}\alpha \psi_1  + \bar{f}(\bz) e^{-\alpha} \psi_2  \label{d9}\\
  \bar{\partial} \psi_2 &= -\half \bar{\partial}\alpha \psi_2  +  \lambda e^{\alpha} \psi_1   \label{d10}
\end{align}
and the same for $\tilde{\psi}_1$, $\tilde{\psi}_2$
 It follows that
\beq
 \psi_1 \simeq \psi_{10}(\theta)\frac{1}{\xi}, \ \ \ \psi_2 \simeq \psi_{20}(\theta)\frac{1}{\xi}, \ \ \ (\xi\rightarrow 0)
  \label{d11}
\eeq
with
\beq
 \frac{\psi_{20}(\theta)}{\psi_{10}(\theta)} = -\lambda e^{-i\theta}
 \label{d11b}
\eeq
 In the case of $\lambda=1$ we can combine this with the reality condition $\psi_2=-i\psi_1^*$ to obtain
\beq
  \frac{\psi_1}{\psi^*_1} \simeq_{\xi\rightarrow 0} i  e^{i\theta}, \ \ \ \ \ \ \ (\lambda=1)
  \label{d12}
\eeq
 In the case of $\lambda=-1$ we obtain 
\beq
 \frac{\phi_1}{\phi^*_1} \simeq_{\xi\rightarrow 0} -i e^{i\theta}, \ \ \ \ \ \ \ (\lambda=-1)
 \label{d13}
\eeq
where we used $\phi_{1,2}$ to denote solutions for $\lambda=-1$. The surface is then described by
\beq
 \X = \left(\begin{array}{cc}\psi_1 & -i\psi^*_1 \\ \tilde{\psi}_1 & -i\tilde{\psi}^*_1 \end{array}\right) \left(\begin{array}{cc}-i\tilde{\phi}^*_1 & i\phi^*_1 \\ -\tilde{\phi}_1 & \phi_1 \end{array}\right)
 = - 2 \mathrm{Im} \left(\begin{array}{cc}\psi^* \tilde{\phi}_1 & \psi_1\phi^*_1 \\ \tilde{\psi}^*_1\tilde{\phi}_1 & \tilde{\psi}_1 \phi_1^* \end{array}\right)
 \label{d14}
\eeq
 The normalization of the solutions should be such that $\det\X=1$. However, when computing the solution in Poincare coordinates the normalization cancels in $x_\pm=X\pm T=\pm\tan\frac{t_\pm}{2}=\pm\tan\frac{t\pm\phi}{2}$:
\beq
 x_+ = \frac{\psi_1 \phi_1^* -\phi_1\psi_1^*}{\tilde{\psi}_1 \phi_1^* -\tilde{\psi}^*_1 \phi_1}, \ \ \ \ \ x_- = \frac{\tilde{\psi}^*_1 \tilde{\phi}_1 -\tilde{\psi}_1 \tilde{\phi}_1^*}{\tilde{\psi}_1 \phi_1^* -\tilde{\psi}^*_1 \phi_1}
 \label{d15}
\eeq
Near the boundary, equations (\ref{d7})-(\ref{d10}) imply that
\beq
 x_+ = \frac{\psi_1}{\tilde{\psi}_1}, \ \ \ \ x_-=-\frac{\tilde{\phi}_1}{\phi_1} , \ \ \ \ (\xi=0)
 \label{d16}
\eeq
 The functions $\psi_1$ and $\tilde{\psi}_1$ are two linearly independent solutions of the linear problem defined in the boundary along $\theta$. It can be obtained from
\beq
 \partial_\theta (\psi_1,\psi_2) = (\psi_1,\psi_2) \tilde{a}_\theta
 \label{d17}
\eeq
 by eliminating $\psi_2$. Defining
\beq
 \chi = \frac{1}{\sqrt{\tilde{a}^\theta_{21}}}
 \label{d18}
\eeq
the equation is 
\beq
 -\partial_\theta^2 \chi(\theta) + V_\lambda(\theta) \chi(\theta) =0
 \label{d19}
\eeq
where
\beq
 V_\lambda(\theta) = -\frac{1}{4} + 6 \beta_2(\theta) - \frac{f}{\lambda} e^{2i\theta} -  \lambda \bar{f} e^{-2i\theta}
 \label{d20}
\eeq
very similar to the Euclidean case. If $\beta_2(\theta)$ and $f(\theta)$ are known, we need to find two linearly independent solution of the equation for $\lambda=1$ to determine $x_+$ as their ratio and
the same for $x_-$ with $\lambda=-1$. Using the result for the Schwarzian derivative of the ratio of two solutions 
\beq
\left\{\frac{\psi_1}{\tilde{\psi}_1},\theta\right\}= -2 V(\theta)
 \label{d21}
\eeq
we find
\beq
 \{ x_\pm(\theta), \theta\} = -2 V_{\lambda=\pm 1}(\theta) =  \frac{1}{2} - 12 \beta_2(\theta) 
                             \pm 2 f e^{2i\theta} \pm 2 \bar{f} e^{-2i\theta} 
 \label{d22}
\eeq
 That means that, if we knew the boundary contour $x_{\pm}(\theta)$ in the conformal parameterization then we could compute $\beta_2(\theta)$
 \beq
  \beta_2(\theta) = \frac{1}{24} \left[1-\{x_+,\theta\} - \{x_-,\theta\} \right]
\label{d22a}
 \eeq
  and also $f(z)$ by using a dispersion relation. As in \cite{WLMA}, one way to find such conformal parameterization is to 
  write eq. (\ref{d19}) after an arbitrary reparameterization $\theta(s)$:
\beq
  -\partial_s^2 \tilde{\chi} + \tilde{V}_\lambda(s) \tilde{\chi}(s) =0
  \label{d221}
\eeq  
with
\begin{align}
\tilde{\chi}(s) &= \frac{1}{\sqrt{\partial_s\theta}}\, \chi(\theta) 
\label{d222}\\
\tilde{V}_\lambda(s) &= (\partial_s\theta)^2\, V_\lambda(\theta(s)) -\half \{\theta(s),s\}
\label{d223}
\end{align}
From eq.(\ref{d22}) it follows that
\beq
 \tilde{V}_{\lambda=\pm1} = -\half \{ x_\pm,s\}
\label{d224}
\eeq
and also, more explicitly,
\begin{align}
\tilde{V}_\lambda(s) &= V_0(s) - \frac{1}{2}\left(\lambda+\frac{1}{\lambda}\right)\, V_1(s) -\frac{i}{2}  \left(\lambda-\frac{1}{\lambda}\right)\, V_2(s)
\label{d225} \\
V_0(s) &= -\frac{1}{4}(\{x_+,s\}+\{x_-,s\}) 
\label{d226}\\
V_1(s) &= \frac{1}{4}(\{x_+,s\}-\{x_-,s\}) = (f\,e^{2i\theta} + \bar{f}\, e^{-2i\theta}) (\partial_s\theta)^2 
\label{d227}\\
V_2(s) &= i\, (f\,e^{2i\theta} - \bar{f}\, e^{-2i\theta}) (\partial_s\theta)^2 
\label{d228}
\end{align}
Thus, knowing the boundary curve $x_\pm(s)$ in an arbitrary parameterization allows the computation of $V_{0,1}(s)$ but leaves $V_2(s)$ undetermined. Similarly as in \cite{WLMA} the real function $V_2(s)$ can be computed by requiring that 
all solutions of the Schr\"odinger equation (\ref{d221}) are anti-periodic in the variable $s$. Once $V_2(s)$ is determined,
it is possible to compute the area and the conformal reparameterization $\theta(s)$. 
  For later use it is convenient to recall the relation to the boundary variables in global coordinates $(t,\phi)$:
\beq
  e^{i(t\pm\phi)}=\frac{1\pm i x_\pm}{1 \mp i x_\pm}, 
\label{d22b}
\eeq

\subsection{Computation of the Area}

To compute the regularized area we used formula (\ref{c21}). It can be simplified by observing that the sinh-Gordon equation (\ref{b10}) implies
\beq
 \partial(\bpartial^2 \alpha- (\bpartial\alpha)^2 ) = - f \bpartial\bar{f} e^{-2\alpha} +4 f\bar{f} \bpartial\alpha e^{-2\alpha}
 \label{d24}
\eeq
Locally, we can rewrite this equation as
\beq
 \partial\left(\frac{2}{\sqrt{\bar{f}}}(\bpartial^2 \alpha- (\bpartial\alpha)^2 )\right) = -4 \bpartial \left(f\sqrt{\bar{f}} e^{-2\alpha} \right)
\label{d25}
\eeq
 If $f$ has no zeros inside the unit disk then this equation defines a conserved current on the world-sheet. At this point
 it is useful to recall that, under a holomorphic coordinate transformation $z\rightarrow w(z)$ the sinh-Gordon equation is invariant provided we change
\begin{align}
 \alpha &\rightarrow \tilde{\alpha} =\alpha - \half \ln \partial w  - \half \ln \bpartial \bar{w} \label{d26a}\\
 f      &\rightarrow \tilde{f} = \frac{f}{(\partial w)^2}
\label{d26b}
\end{align}
in particular implying
\beq
 \sqrt{f} dz = \sqrt{\tilde{f}} dw, 
\label{d27}
\eeq
namely $\chi = \sqrt{f}\, dz$ is a holomorphic 1-form and then
\beq
 W(z) = \int^z \chi = \int^z \sqrt{f(z')}\, dz'
\label{d28}
\eeq
is a function (0-form) on the disk such that $\chi=dW$. On the other hand
\beq
 2 \left[\bpartial_{\bar{w}}^2 \tilde{\alpha} - (\bpartial_{\bar{w}}\tilde{\alpha})^2\right] = 
 \frac{1}{(\bpartial\bar{w})^2} \left\{2 \left[\bpartial^2\alpha-(\bpartial\alpha)^2 \right]-\{\bar{w},\bar{z}\}\right\}
\label{d29}
\eeq
namely $2[\bpartial^2\alpha-(\bpartial\alpha)^2]$ transforms as a Schwarzian derivative. Since the difference between two Schwarzian derivatives transforms homogeneously, we can rewrite eq.(\ref{d25}) as the conservation of the current 
\begin{align}
 j &= j_z dz + j_{\bar{z}} d\bar{z} 
 \label{d30a} \\
 j_z &= -4 f \sqrt{\bar{f}} e^{-2\alpha}
 \label{d30b} \\
 j_{\bar{z}} &= \frac{2}{\sqrt{\bar{f}}} [\bpartial^2\alpha-(\bpartial\alpha)^2] -  \frac{1}{\sqrt{\bar{f}}} \, \{\bar{W},\bar{z}\}
 \label{d30c} \\
 dj &= 0
 \label{d30d}
\end{align}
 where we used the function $W(z)$ defined in eq.(\ref{d28}) to write a current that transforms appropriately under a coordinate transformation. Otherwise the extra term $-  \frac{1}{\sqrt{\bar{f}}} \, \{\bar{W},\bar{z}\} $ does not play any role since it is anti-holomorphic. Finally, we follow \cite{AMSV} and write the area as ($d\sigma\wedge d\tau= \frac{i}{2} dz\wedge d\bar{z}$)
\begin{align}
 \cA_f +2 \pi &=  4 \int_D f\bar{f} e^{-2\alpha} d\sigma d\tau = -\frac{i}{2} \int_D j\wedge \bar{\chi} \\
       & =  - \frac{i}{2} \int_D j\wedge d \bar{W} =\frac{i}{2}  \int_D d (\bar{W} j) 
\label{d31}
\end{align}
The integral is over the unit disk whose boundary is parameterized as $z=e^{i\theta}$. Integrating by parts we find
\beq
 \cA_f = - 2\pi + \frac{i}{2} \oint_{\partial D} \bar{W} (j_z \partial_\theta z+ j_{\bar{z}} \partial_\theta\bar{z})\, d\theta
\label{d32}
\eeq
At the boundary $\alpha$ diverges and then, from eq.(\ref{d30b}) $j_z$ vanishes whereas, from eq.(\ref{d2})
\beq
 j_{\bar{z}} = \frac{1}{\sqrt{\bar{f}}} \left(12\beta_2(\theta) e^{2i\theta} - \{\bar{W},\bar{z}\} \right) 
 + \cO(\xi) \ \ \ \ \ \ (\xi=1-r^2\rightarrow 0)
\label{d33}
\eeq 
Thus
\beq
  \cA_f = - 2\pi + \frac{i}{2} \oint_{\partial D} \frac{\bar{W}}{\sqrt{\bar{f}}} \left(12\beta_2(\theta) - \{\bar{W},\bar{z}\} \right)\partial_\theta\bar{z}\, d\theta
\label{d34}
\eeq
Using eq.(\ref{d22a}) together with the simple result 
\beq
 \{x_\pm,\theta\} = \half - e^{-2i\theta} \{x_\pm,\bar{z}\}, \ \ \ \ \bar{z} = e^{-i\theta}
\label{d35}
\eeq
it follows that
\beq
 \cA_f = - 2\pi + \frac{i}{2} \oint_{\partial D} \frac{\bar{W}}{\bpartial\bar{W}} \left[\half \{x_+,\bar{z}\}+\half\{x_-,\bar{z}\} - \{\bar{W},\bar{z}\} \right]\ d\bar{z}
\label{d36}
\eeq
This result is invariant under reparameterizations of the boundary and therefore we can choose an arbitrary parameter $s$ instead of $\bar{z}$:
\beq
 \cA_f = -2\pi + \frac{i}{2} \oint \frac{\bar{W}}{\partial_s \bar{W}} \left[\half\{x_+,s\}+\half\{x_-,s\} - \{\bar{W},s\}\right] \, ds
 \label{d23}
\eeq
 Finally inside the disk we can take any other conformal parameterizations. In the next section we use $W(z)$ as a coordinate and just denote it as $z$. In that case the function $f(z)=1$ and the boundary of the world-sheet is given
 by a curve $z(s)$ that has to be found as part of the solution. 

\section{Solutions in terms of theta functions}

 In this section we discuss exact analytical solutions to the minimal area surface problem that can be written in terms of Riemann Theta functions. It follows along the lines of similar solutions constructed in \cite{IKZ,KZ,IK}.
We are going to consider the case where the analytic function $f(z)$ appearing in eq.(\ref{b10}) has no zeros inside the unit circle and therefore can be set to $f(z)=1$ by an appropriate conformal transformation of the unit circle into a new domain 
in the complex plane that has to be found as part of the solution. The equation for $\alpha$ reduces to the sinh-Gordon equation
\beq
 \partial \bar{\partial} \alpha = 2\sinh 2\alpha
 \label{e1}
\eeq
that has known solutions in terms of Riemann Theta functions associated to hyperelliptic Riemann surfaces. We are going to define such a surface by an equation in $\mathbb{C}^2$ 
\beq
 \mu^2 = \prod_{i=1}^{2g+1} (\lambda-\lambda_i)
 \label{e2}
\eeq
where $g$ is the (arbitrary) genus and $(\mu,\lambda)$ parameterize $\mathbb{C}^2$.  
 For the solution to be real the branch points have to be symmetric under the involution $T:\lambda \rightarrow 1/\bar{\lambda}$, see also eq.(\ref{b4}).  We should then choose a basis of cycles $\{a_i,b_i\}$ such that the involution  maps:
\beq
(Ta)_i = T_{ij} a_j, \ \ (Tb)_i = -T_{ij} b_j. 
\label{e2b}
\eeq
This choice defines the $g\times g$ matrices
\beq
 C_{ij} = \oint_{a_i} \frac{\lambda^{j-1}}{\mu(\lambda)}\, d\lambda , \ \ 
 \tilde{C}_{ij} = \oint_{b_i} \frac{\lambda^{j-1}}{\mu(\lambda)}\, d\lambda 
 \label{e3}
\eeq
as well as a basis of holomorphic differentials
\beq
 \omega_i =  \sum_{j=1}^g  \frac{\lambda^{j-1}}{\mu(\lambda)} C^{-1}_{ji}
 \label{e4}
\eeq
such that
\beq
 \oint_{a_i} \omega_j = \delta_{ij}, \ \ \ \ \oint_{b_i} \omega_j = \Pi_{ij} 
 \label{e5}
\eeq
 where $\Pi=\tilde{C}C^{-1}$ is the period matrix of the Riemann surface. The next step is to choose two branch points $p_{1,3}\equiv(\lambda_{1,3},\mu=0)$ that map into each other under the involution $T$. In addition we require that the path connecting them is an even half-period: $\mathcal{C}_{13}=\half(\Delta_{2i} a_i + \Delta_{1i} b_i)$, with $\Delta_1^t\Delta_2$ and even integer. 
 This half period define a Theta function with characteristics that we call 
\beq
\hat{\theta}(\zeta) = \theta\left[\begin{array}{c}\Delta_1 \\ \Delta_2 \end{array}\right](\zeta), \ \ \ \ \zeta\in \mathbb{C}^g
 \label{e6}
\eeq
 Using the properties under the involution  $T:\lambda \rightarrow 1/\bar{\lambda}$ it is easy to prove that
\begin{align}
 C_{ij} &= - e^{-i\phi}\ T_{il} C^*_{l\,g-j+1}    \label{e7}\\
 \tilde{C}_{ij} &= e^{-i\phi}\ T_{il} \tilde{C}^*_{l\,g-j+1}  \label{e8}\\
  \Pi^* &= -T \Pi T   \label{e1}
\end{align}
where $\phi$ is defined through
\beq
 e^{2i\phi} = \prod_{i=1}^{2g} \lambda_i  \label{e9}
\eeq
These results imply that, if $\zeta^*=\pm T\zeta$, then $\theta(\zeta),\thetah(\zeta) \in \, \mathbb{R}$.
As we approach the branch points $p_{1,3}$, the vector of holomorphic differentials $\omega(\lambda)$ diverges as $1/\mu(\lambda)$; for that reason it is convenient to define a new vector 
\beq
\omega_f(\lambda)_i = \sum_{j=1}^g \lambda^{j-1} C_{ji}^{-1} = \mu(\lambda)\, \omega_i
 \label{e10}
\eeq
 and two particular values:
\beq
 \omega_1 = -\frac{1}{\lambda_1^{g-1}}\omega_f(\lambda_1), \ \ \  \omega_3 = \omega_f(\lambda_3)
 \label{e11}
\eeq
 where $\lambda_{1,3}$ are the projections of the points $p_{1,3}$. If we further define the constant
\beq
 C_\pm^2 = -\frac{\thetah^2(a)}{D_1\theta(a)D_3\theta(a)}
 \label{e12}
\eeq
then we obtain that the following reality condition is satisfied:
\beq
 (C_\pm \omega_1)^* = T (C_\pm \omega_3) 
 \label{e13}
\eeq
Under all these conditions, from eq.(\ref{D13idC}) in the appendix, it follows that a real solution to the sinh-Gordon 
equation can be written as
\beq
 e^{\alpha} = C_{\alpha} \frac{\theta(\zeta)}{\hat{\theta}(\zeta)}, \ \ \ \ \ \zeta= C_\pm (\omega_1 z + \omega_3 \bz), \ 
 \label{e14}
\eeq
where $C_{\alpha}$ is a constant equal to $\pm1$, chosen so that $e^\alpha$ is positive in the region of interest. Such region of interest is taken to be a connected domain in the complex plane bounded by a curve where $\hat{\theta}$ vanishes, namely $\alpha$ diverges. It should be noted that the condition that
$\thetah$ vanishes is only one real equation since $\thetah$ is real, a general theta function with arbitrary characteristics would be complex and the condition that it vanishes would only be satisfied at isolated points in the world-sheet. 
 
 The next step is to solve the linear problem for $\Psi$, namely eq.(\ref{b12}). To this end we choose an arbitrary point $p_4$ on the Riemann surface, for example on the upper sheet, and write the solutions as
\begin{align}
 \psi_1          &=  e^{\mu_4 z + \nu_4 \bar{z}}                                        e^{\frac{\alpha}{2}} \frac{\theta(\zeta+\lv)}{\theta(\zeta)}  \label{e15a} \\
 \psi_2          &= A e^{\mu_4 z + \nu_4 \bar{z}} \frac{\theta(a-\lv)}{\thetah(a-\lv)}   e^{\frac{\alpha}{2}} \frac{\hat{\theta}(\zeta+\lv)}{\theta(\zeta)}  \label{e15}
\end{align}
where
\begin{align}
 \mu_4 &= - C_{\pm} D_1\ln\frac{\thetah(a)}{\theta(a-\int_1^{4})}   \label{e16} \\
 \nu_4 &= - C_{\pm} D_3\ln\frac{\thetah(a)}{\thetah(a-\int_1^{4})}  \label{e17}
\end{align}
and the constant $A$ is given by
\beq
  A = -C_\pm \frac{D_3\theta(a)}{\thetah(a)}  \label{e18}
\eeq
 It is straight forward to use the properties (\ref{firstder}) of the theta functions to prove that $\psi_{1,2}$ solve the linear equations (\ref{d7})--(\ref{d10}) with a spectral parameter
\beq
 \lambda = \frac{D_3\theta(a)}{D_1\theta(a)} \left( \frac{\theta(a-\int_1^4)}{\thetah(a-\int_1^4)} \right)^2
 \label{e19}
\eeq
Recall that real solutions require $|\lambda|=1$ (see eq.(\ref{c18b})) which restricts the possible points $p_4$ 
that can be chosen, in fact, as discussed in the appendix, we have to choose $|\lambda_4|=1$.

It is easy to see that $|\lambda|=1$ implies that, if $(\psi_1,\psi_2)$ is a solution to eqs. (\ref{d7}), (\ref{d8}) then so is $(\psi_2^*,\psi_1^*)$. However, one can check that for the solutions in eqs.(\ref{e15a}), (\ref{e15}) such solution is the same as the original (up to an overall constant). Instead, another, linearly independent solution, to equations (\ref{d7})--(\ref{d10}) is obtained by choosing the corresponding point on the lower sheet of the Riemann surface that we denote as $p_{\bar{4}}$. 
Since $p_1$ is a branch point we have $\int_1^4  = - \int_1^{\bar{4}}$. The value of the spectral parameter does not change since it can be seen that
\beq
 \lambda = \frac{D_3\theta(a)}{D_1\theta(a)} \left( \frac{\theta(a+\int_1^4)}{\thetah(a+\int_1^4)} \right)^2
 \label{e20}
\eeq
 Finally, we also need to find solutions with spectral parameter $-\lambda$. For that purpose we choose a point $p_5$ on the upper sheet of the Riemann surface such that
\beq
 - \lambda = \frac{D_3\theta(a)}{D_1\theta(a)} \left( \frac{\theta(a-\int_1^5)}{\thetah(a-\int_1^5)} \right)^2
 \label{e21}
\eeq
 and the corresponding point $p_{\bar{5}}$ on the lower sheet. It might seem that it is difficult to find such point but it is actually quite simple as explained in the particular examples given later in the paper where it is also shown how to find $p_{4}$ such that $|\lambda|=1$. 
 
At this point we can write a complete solution to the linear problem as
\begin{align}
 \phi_{11}          &=  e^{\mu_4 z + \nu_4 \bar{z}}                                        e^{\frac{\alpha}{2}} \frac{\theta(\zeta+\lv)}{\theta(\zeta)}  \label{e22}\\
 \phi_{21}          &= A e^{\mu_4 z + \nu_4 \bar{z}} \frac{\theta(a-\lv)}{\thetah(a-\lv)}   e^{\frac{\alpha}{2}} \frac{\hat{\theta}(\zeta+\lv)}{\theta(\zeta)}  \label{e23} \\
 \tilde{\phi}_{11}  &=  e^{-\mu_4 z - \nu_4 \bar{z}}   e^{\frac{\alpha}{2}} \frac{\theta(\zeta-\lv)}{\theta(\zeta)}  \label{e24}\\
 \tilde{\phi}_{21}  &= A e^{-\mu_4 z - \nu_4 \bar{z}} \frac{\theta(a+\lv)}{\thetah(a+\lv)}   e^{\frac{\alpha}{2}} \frac{\thetah(\zeta-\lv)}{\theta(\zeta)}  \label{e25}
\end{align}
\begin{align}
 \phi_{12}          &=  e^{\mu_5 z + \nu_5 \bar{z}}                                        e^{\frac{\alpha}{2}} \frac{\theta(\zeta+\lvt)}{\theta(\zeta)}  \label{e26} \\
 \phi_{22}          &= A e^{\mu_5 z + \nu_5 \bar{z}} \frac{\theta(a-\lvt)}{\thetah(a-\lvt)}   e^{\frac{\alpha}{2}} \frac{\hat{\theta}(\zeta+\lvt)}{\theta(\zeta)} \label{e27} \\
 \tilde{\phi}_{12}  &=  e^{-\mu_5 z - \nu_5 \bar{z}}   e^{\frac{\alpha}{2}} \frac{\theta(\zeta-\lvt)}{\theta(\zeta)}  \label{e28}\\
 \tilde{\phi}_{22}  &= A e^{-\mu_5 z - \nu_5 \bar{z}} \frac{\theta(a+\lvt)}{\thetah(a+\lvt)}   e^{\frac{\alpha}{2}} \frac{\thetah(\zeta-\lvt)}{\theta(\zeta)}   \label{e29}
\end{align}
where the constant $A$ was defined in eq.(\ref{e18}). Using these functions we can write a solution $\Psi$ to eq.(\ref{b12}):
\begin{align}
 \Psi(\lambda) &= \left(\begin{array}{cc}\phi_{11} & \phi_{21}\\ \tilde{\phi}_{11} & \tilde{\phi}_{21} \end{array}\right),  \label{e31}\\
 \Psi(-\lambda) &= \left(\begin{array}{cc}\tilde{\phi}_{12} & \tilde{\phi}_{22} \\ \phi_{12} & \phi_{22}  \end{array}\right),  \label{e31}
\end{align}
This is not the whole story since the actual matrices $\Psi$ also have to satisfy the reality conditions (\ref{c18}). Fortunately this problem is easily solved by first defining the linear combinations
\begin{align}
  \Psi_{F}(\lambda)  &= \Psi(\lambda) + \sigma_1 [\Psi(1/\bar{\lambda})]^* \sigma_1  \label{e32}\\
  \Psi_{F}(-\lambda) &= \Psi(-\lambda) + \sigma_1 [\Psi(-1/\bar{\lambda})]^* \sigma_1   \label{e33}
\end{align}
that satisfy the same equations due to the symmetry (\ref{c17}) of the flat current but in addition satisfy the reality condition
\beq
 [\Psi_F(\lambda)]^* = \sigma_1 \Psi_F(1/\bar{\lambda}) \sigma_1
\eeq
Then we define 
\beq
 \Psi_R(\lambda)= i \hat{R} \Psi_F
\eeq
 that satisfy the reality condition (\ref{c18}) as required and can be checked using the definition of $\hat{R}$ in eq.(\ref{c16}). 
 Finally we can write the solution to the non-linear problem as
\begin{align}
 \mathbb{X}_0 &= \Psi_{R}(\lambda)\Psi_{R}(-\lambda)^{-1} \label{e34} \\
 \mathbb{X}  &= \frac{1}{\sqrt{\det \mathbb{X}_0}} \mathbb{X}_0 \label{e35}
\end{align}
The intermediate matrices $\Psi_F(\lambda)$ are useful since we can equally well write the solution in the form
\begin{align}
\mathbb{X}_F = \Psi_F(\lambda)\Psi_F(-\lambda)^{-1} &= \left(\begin{array}{cc}X_{-1}+iX_0 & X_1+iX_2 \\ X_1-iX_2 & X_{-1}-iX_0 \end{array}\right) \\ &= \left(\begin{array}{cc}\cosh\rho\ e^{it} & \sinh\rho\ e^{i\phi} \\ \sinh\rho\ e^{-i\phi}  & \cosh\rho\ e^{-it} \end{array}\right)  \label{e36}
\end{align}
This gives the shape of the surface analytically. In the next section we give particular examples to get an idea of the shape of these solutions.  

\subsection{Computation of the area}

 The regularized area can be computed by using the formula (\ref{c21})
\beq
\cA_f = -2\pi + 4 \int_D e^{-2\alpha} dzd\bz
\label{f1}
\eeq
where we set $f(z)=1$ since we are considering that case. The domain $D$ is the region of the complex plane bounded by the curve where $\thetah$ vanishes. Furthermore, from eqs.(\ref{D13id}) and (\ref{e12}) we find 
\beq
 e^{-2\alpha} = \frac{1}{C_\alpha^2}\frac{\thetah^2(\zeta)}{\theta^2(\zeta)} = \frac{C_\pm^2}{C_\alpha^2} D_{13}\ln \frac{\thetah(a)}{\theta(\zeta)} = -\frac{1}{C_\alpha^2} \partial\bar{\partial}\ln \theta(\zeta) + \frac{C_\pm^2}{C_\alpha^2} D_{13}\ln\thetah(a)
 \label{f2}
\eeq
Thus, the regularized area is equal to
\beq
 \cA_f = -2\pi + 4 \frac{C_\pm^2}{C_\alpha^2}\, D_{13}\ln\thetah(a)\, \cA_{WS} -\frac{4}{C_\alpha^2} \int  \partial\bar{\partial}\ln \theta(\zeta)\, dzd\bz
\label{f3}
\eeq 
 where $ \cA_{WS}$ is the world-sheet area, namely the area of the domain $D$ of the complex plane that maps to the minimal surface. The last integral can be done using Gauss' theorem in the form
\beq
\int \partial\bar{\partial} F dzd\bz = -\frac{i}{4} \oint (\partial F dz-\bar{\partial}F d\bz) = -\frac{i}{2}\oint \partial F \partial_\theta z d\theta
\label{f4}
\eeq
where in the last equality we used that $\oint (\partial F dz+\bar{\partial}F d\bz) = \oint dF =0$. The final result for the Area is then
\beq
 \cA_f = -2\pi + 4 \frac{C_\pm^2}{C_\alpha^2}\, D_{13}\ln\thetah(a)\, \cA_{WS} + \frac{2i}{C_\alpha^2} \oint \partial \ln\theta(\zeta)\, \partial_\theta z\, d\theta
\label{f5}
\eeq
with $\partial \ln\theta(\zeta)= C_\pm D_1\ln \theta(\zeta)$ evaluated along the boundary. This gives a practical way to evaluate the area for the solutions discussed in this section. We can now verify eq.(\ref{d23}).
Indeed starting from (\ref{d23}) and using eqs.(\ref{Sd12}), (\ref{Sd14}) we obtain
\begin{align}
\cA_f &=  -2\pi + \frac{i}{2C_\alpha^2}
 \oint \frac{\bar{z}\, ds}{\partial_s\bar{z}} \left(\half\{x_+,s\} +\half\{x_-,s\} - \{\bar{z},s\} \right) 
 \label{f6} \\
 &= -2\pi + 2i \frac{C_{\pm}^2}{C_\alpha^2} \oint \bar{z} \partial_s \bar{z}\ D_3^2\ln \theta(\zeta_s)\ ds
\label{f7}
\end{align}
where we renamed $W\rightarrow z$ for simplicity since we use $W$ as the world--sheet coordinate. Furthermore, since
\beq
 \partial_s D_3 \ln \theta(\zeta_s) = C_\pm \partial_s \bar{z} \, D_3^2\ln\theta(\zeta_s) + C_\pm \partial_s z\, D_{13} \ln\theta(\zeta_s)
\label{f8}
\eeq
and also from eq.(\ref{bid5}) we find
\beq
\cA_f = -2\pi  -2 i  \frac{C_\pm}{C_\alpha^2} \oint \partial_s \bar{z} \ D_3 \ln\theta(\zeta_s) \ ds- 2i \frac{C^2_\pm}{C_\alpha^2} D_{13} \ln \thetah(a) \oint \bar{z} \partial_s z\ ds
\label{f9}
\eeq
Finally, since the world-sheet area $\cA_{WS}$ is given by 
\beq
 \cA_{WS} = -\frac{i}{2} \oint \bar{z} \partial_s z \ ds 
\label{f10}
\eeq
 and we can integrate by parts 
\beq
\oint \partial_s\bar{z}\ D_3 \ln\theta(\zeta_s) \ ds = - \oint \partial_s z \  D_1 \ln\theta(\zeta_s) \ ds
\label{f11}
\eeq
we find
\beq
\cA_f = -2\pi  + 2 i  \frac{C_\pm}{C_\alpha^2} \oint \partial_s z\ D_1 \ln\theta(\zeta_s) \ ds + 4 \frac{C^2_\pm}{C_\alpha^2} D_{13} \ln \thetah(a)  \cA_{WS}
\label{f12}
\eeq
in perfect agreement with eq.(\ref{f12}).

 \subsection{Boundary curve}
 
 The boundary curve associated with these minimal area surfaces can be derived by using eqns.(\ref{e36}), and (\ref{e31})-(\ref{e33})
\beq
 e^{i(t+\phi)} = \left. \frac{(\mathbb{X}_F)_{12}}{(\mathbb{X}_F)_{22}} \right|_{\mathrm{bdry.}} = \left.
 \frac{-(\Psi_F(\lambda))_{11}(\Psi_F(-\lambda))_{12}+(\Psi_F(\lambda))_{12}(\Psi_F(-\lambda))_{11}}{-(\Psi_F(\lambda))_{21}(\Psi_F(-\lambda))_{12}+(\Psi_F(\lambda))_{22}(\Psi_F(-\lambda))_{11}} \right|_{\mathrm{bdry.}} 
 \label{g1}
\eeq
 and similarly for $\hat{x}_-$. This can be greatly simplified by studying the behavior of the functions near the boundary as in eq.(\ref{d11}). However, since we are using here a world-sheet parameterization such that $f(z)=1$, the world-sheet is bounded by a curve $z(s)$ which generically is not a circle. For that reason we revisit the derivation. Consider a point $z_0$ at the world-sheet boundary and expand the coordinate $z$ as
\beq
 z \simeq z_0 + (s+i\xi)\, \partial_s z(s) 
  \label{g2}
\eeq
 where $s$ represents fluctuations along the boundary and $\xi$ towards the inside of the world-sheet ($\xi=0$ is the boundary). Instead of eq.(\ref{d11}) we now find
\beq
 \psi_1 \simeq \psi_{10}(s) e^{-|\partial_sz|\ln \xi}, \ \   \psi_2 \simeq \psi_{20}(s) e^{-|\partial_sz|\ln \xi}, \ \ \ \ (\xi\rightarrow 0)
  \label{g3}
\eeq
with 
\beq
 \frac{\psi_{10}(s)}{\psi_{20}(s)} = -\frac{i}{\lambda} \frac{\partial_s z}{|\partial_s z|}
 \label{g4}
\eeq
 Since $\psi_{1,2}$ obey the same equations as $\psi_{1,2}$ but with $\lambda \leftrightarrow -\lambda$ it follows that
$\phi_{1,2}$ behave in the same way with
\beq
  \frac{\phi_{10}(s)}{\phi_{20}(s)} = \frac{i}{\lambda} \frac{\partial_s z}{|\partial_s z|}
 \label{g5}
\eeq
 We can now simplify (\ref{g1}) to
\beq
e^{i(t+\phi)} = \left. \frac{\phi_{21}+\tilde{\phi}_{11}^*}{\tilde{\phi}_{21}+\phi_{11}^*} \right|_{\mathrm{bdry.}}
 \label{g6}
\eeq
As mentioned before, in this case $(\phi_{11},\phi_{21})$ and $(\phi^*_{21},\phi^*_{11})$ are linearly dependent solutions
implying that $\phi^*_{11}/\phi_{21}$ is constant on the world-sheet. In fact, using that
\beq
 T \left(\lv\right)^* = \lv - \int_1^3
 \label{g7}
\eeq
where the matrix $T$ is defined in eq.(\ref{e2b}), we obtain
\begin{align}
 \mu_4^*-\nu_4 &= \mu_5^*-\nu_5 = -i\pi C_\pm\, (\Delta_1^t.\omega_3)
  \label{g8} \\
 \mu_4-\nu_4^* &= \mu_5-\nu_5^* =  i\pi C_\pm\, (\Delta_1^t.\omega_1)
    \label{g8b} \\
 \theta^*(\zeta+\lv) &= e^{i\pi\Delta_1^t(\zeta+\lv)+\half i\pi\Delta_1^t\Delta_2-\frac{1}{4}i\pi\Delta_1^t\Pi\Delta_1} \thetah(\zeta+\lv)
 \label{g9}
\end{align}
Now the constant can be computed explicitly as
\beq
 B_1=  \frac{\tilde{\phi}_{11}^*}{\tilde{\phi}_{21}}  = -\frac{\phi_{11}^*}{\phi_{21}}  = \frac{\hat{\theta}(a+\lv)}{\theta(a+\lv)}\frac{1}{A}\ e^{i\pi\Delta_1^t \lv + \frac{i\pi}{2}\Delta_1^t\Delta_2-\frac{i\pi}{4}\Delta_1^t\Pi\Delta_1}
 \label{g10}
\eeq
where $A$ was defined in eq.(\ref{e18}). Finally we obtain
\beq
 e^{it_+} = \frac{1+B_1 \hat{x}_+}{-B_1 + \hat{x}_+}
 \label{g11}
\eeq
where
\beq
 \hat{x}_+ = \frac{\tilde{\phi}_{21}}{\phi_{21}} = - e^{-2\mu_4 z-2\nu_4\bz}\, \frac{\hat{\theta}(\zeta-\lv)}{\hat{\theta}(\zeta+\lv)}
 \label{g12}
\eeq
Similarly
\beq
 e^{it_-} = \frac{1-B_2 \hat{x}_-}{B_2 + \hat{x}_-}
 \label{g13}
\eeq
where
\beq
 \hat{x}_-  = e^{-2\tilde{\mu}_+ \bar{z}-2\tilde{\mu}_-z}\, \frac{\hat{\theta}(\zeta-\lvt)}{\hat{\theta}(\zeta+\lvt)}
 \label{g14}
\eeq
and
\beq
B_2= \frac{\hat{\theta}(a+\lvt)}{\theta(a+\lvt)}\frac{1}{A}\ e^{i\pi\Delta_1^t \lvt + \frac{i\pi}{2}\Delta_1^t\Delta_2-\frac{i\pi}{4}\Delta_1^t\Pi\Delta_1}
 \label{g15}
\eeq
It is important to note that $\hat{x}_\pm$ and $x_{\pm}$ are related by an
 $SL(2,\mathbb{C})$ transformation (as follows from eqs.(\ref{d22b}), (\ref{g11}))  implying that
\beq
 \{\hat{x}_\pm,s\} = \{x_\pm,s\}
 \label{g16}
\eeq
Namely $\hat{x}_{\pm}(s)$ is a conformally equivalent (but generally complex) description of the Wilson loop 

\section{Examples}

 To illustrate the solutions we describe two Wilson loops associated with genus $g=2$ auxiliary surfaces. These examples make clear the shape of the solutions
we are discussing and also provide the reader with concrete numbers that s/he can reproduce and use as a basis for further work. For the same reason the results are rounded to just a few significant figures. \footnote{These calculations can be easily done using Maple or Mathematica. 
}

\subsection{Example 1}
\begin{figure}
  \centering
  \includegraphics[width=10cm]{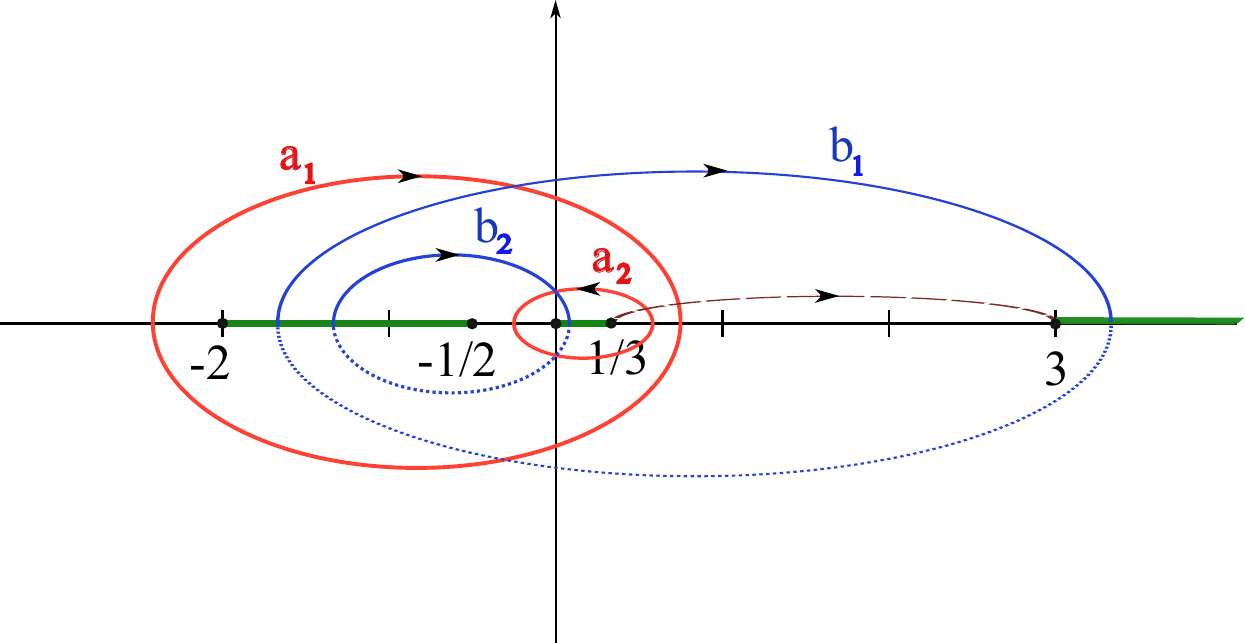}
  \caption{Hyperelliptic Riemann surface for Example 1. Cuts are in green, the brown path is a half period used to define $\hat{\theta}$. }
  \label{cuts1}
  \end{figure}

\begin{figure}
  \centering
  \includegraphics[width=15cm]{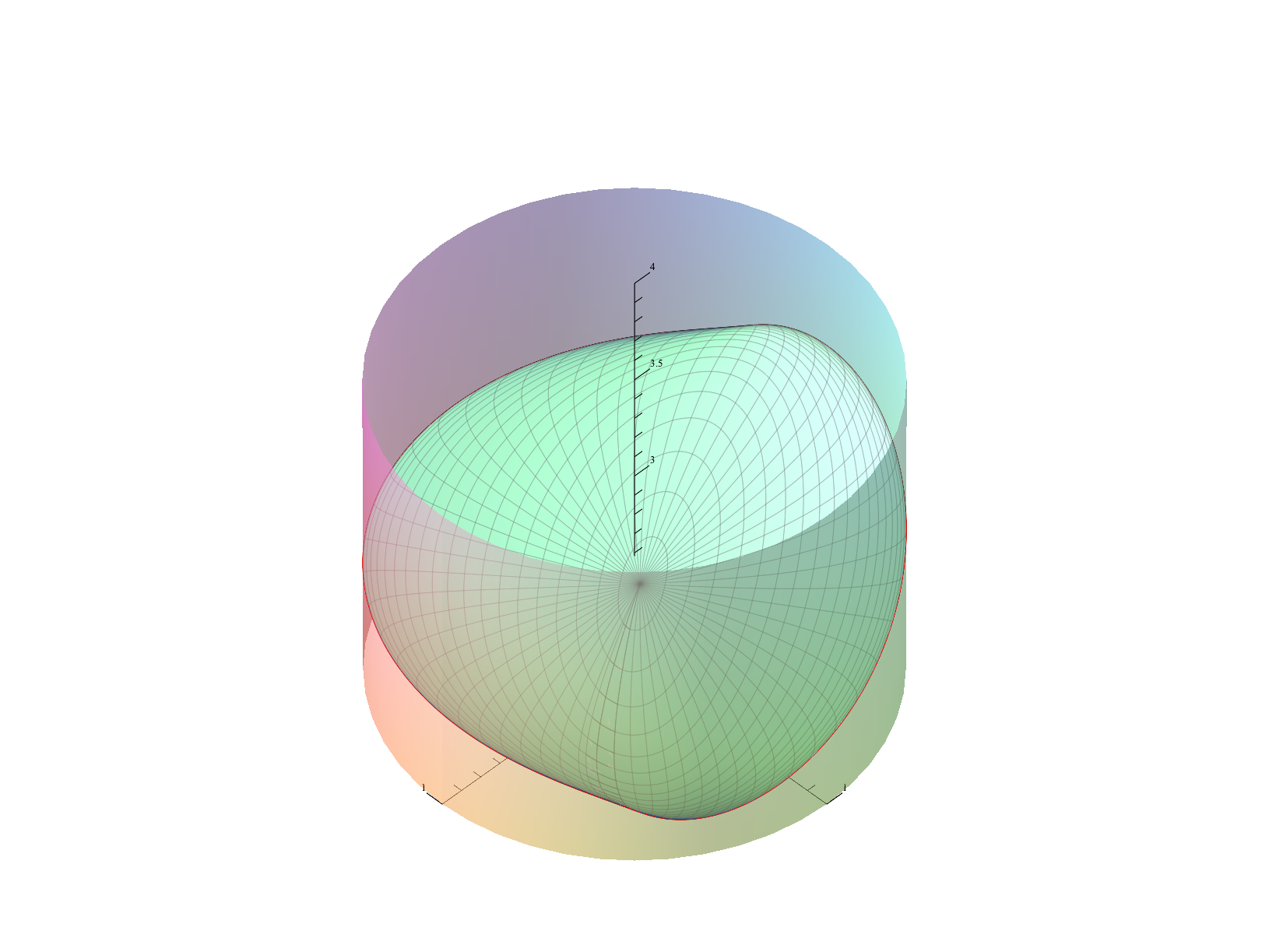}
  \caption{Minimal area surface embedded in \ads{3} in global coordinates $[t,\rho,\phi]$. The vertical direction is time $t$, the radial direction is $\tanh \rho$ and the angle is $\phi$.}
  \label{surface1}
  \end{figure}

In example 1, we choose a surface with branch points $-2,-\half,0,\frac{1}{3},3,\infty$. This surface has the required invariance under $\lambda \leftrightarrow 1/\bar{\lambda}$. In addition it also has the symmetry $\lambda \leftrightarrow 1/\lambda$ that plays no role in the construction but simplifies the calculations\footnote{For example the hyperelliptic integrals appearing in the period matrix can be reduced to ordinary elliptic integrals through the change of variables $\lambda=(1+\sqrt{u})/(1-\sqrt{u})$.}. We choose a basis of cycles as depicted in figure \ref{cuts1} such that property (\ref{e2b}) is satisfied. The period matrix is then 
\beq
 \Pi = \left(\begin{array}{cc} 1.2063\, i & 0.4441\, i \\ 0.4441\, i & 1.2063\, i  \end{array}\right)
 \label{h1}
\eeq
A zero $a$ of the $\theta$ function can be found by choosing an arbitrary odd period, for example 
\beq
 a = \half (\mathbb{I} + \Pi) \left(\begin{array}{cc}0 \\ 1\end{array}\right)
\eeq
 Now we choose two branch points $p_{1,3}$ such that the half-period $\mathcal{C}_{13}=\half(\Delta_2+\Pi \Delta_1)$ connecting them is even. We select $p_1=\frac{1}{3}$, $p_3=3$
 and thus
\beq
\Delta_1 = \left(\begin{array}{c} 1 \\ -1 \end{array} \right)  \ \ \  \Delta_2 = \left(\begin{array}{c} 0 \\ 0 \end{array} \right)  
\eeq
 which, from eq.(\ref{e6}), define $\thetah$.
 Furthermore, the vectors $\omega_{1,3}$ in eq.(\ref{e11}) follow from eq.(\ref{e10}) as
\beq
 \omega_1 = -3 \omega_f(\frac{1}{3})=\left(\begin{array}{cc} 0.1738\\-0.9256 \end{array}\right), \ \ \omega_3 = \omega_f(3)= \left(\begin{array}{cc}-0.9256\\ 0.1738\end{array}\right)
\eeq
 and a solution of the sinh-Gordon equation can then be written as
\beq
 e^{\alpha} = \frac{\theta(\zeta)}{\thetah(\zeta)}
\eeq
with (see eq.(\ref{e14}))
\beq
 \zeta = \left(\begin{array}{c} -0.4425\, i\, \bar{z}+ 0.0831\, i\, z \\ 0.0831\, i\, \bar{z}- 0.4425\, i\, z \end{array}\right) 
\eeq
using 
\beq
C_\pm^2 =  \frac{\thetah^2(a)}{D_1\theta(a)D_3\theta(a)} =-\frac{8}{35}. 
\eeq
as follows form eq.(\ref{pid4}).
 Now we choose two points $p_{4,5}$ on the Riemann that determine the values of the spectral parameter $\lambda$ through eq.(\ref{e19}) that, following the results in appendix A.2, can be inverted to give
\beq
 \lambda_4(\gamma) =  \frac{1+3e^{i\gamma} }{3+e^{i\gamma}}, \ \ \ \ \lambda=e^{i\gamma}
\eeq
where we emphasized that the spectral parameter $\lambda$ has to have modulus one. For this example we choose
\beq
  \lambda_4 = \lambda(0.1), \ \ \lambda_5 = \lambda(\pi+0.1)
\eeq
that determine the points $p_{4,5}$ in the upper sheet and $p_{\bar{4},\bar{5}}$ in the lower sheet. Now we can compute
\begin{align}
 \lv &=  \left(\begin{array}{c} -0.0056 +.1053 i \\ 0.0056 -.2758 i \end{array}\right) \\
 \lvt &= \left(\begin{array}{c} .2091-.1767 i\\ -.2091-.5578 i \end{array}\right) \\ 
 \nu_4 &=   -.3278-0.0259 i\\
 \mu_4 &=  -1.9791+0.0259 i \\
 \nu_5 &= .2533+.1282 i \\
 \mu_5 &= -1.3980-.1282  i 
\end{align} 
which allows us to plot the surface as seen in figure \ref{surface1}. The boundary curve can be obtained from the limit near the boundary or equivalently using equations (\ref{g11})-(\ref{g14}) with 
\begin{align}
 B_1 &= -.9964+0.0852 i \\
 B_2 &= .9532-.3022 i \\
\end{align}
Finally the regularized area can be found to be
\beq
 A_f = -5.876
\eeq

\subsection{Example 2}
 
\begin{figure}
  \centering
  \includegraphics[width=10cm]{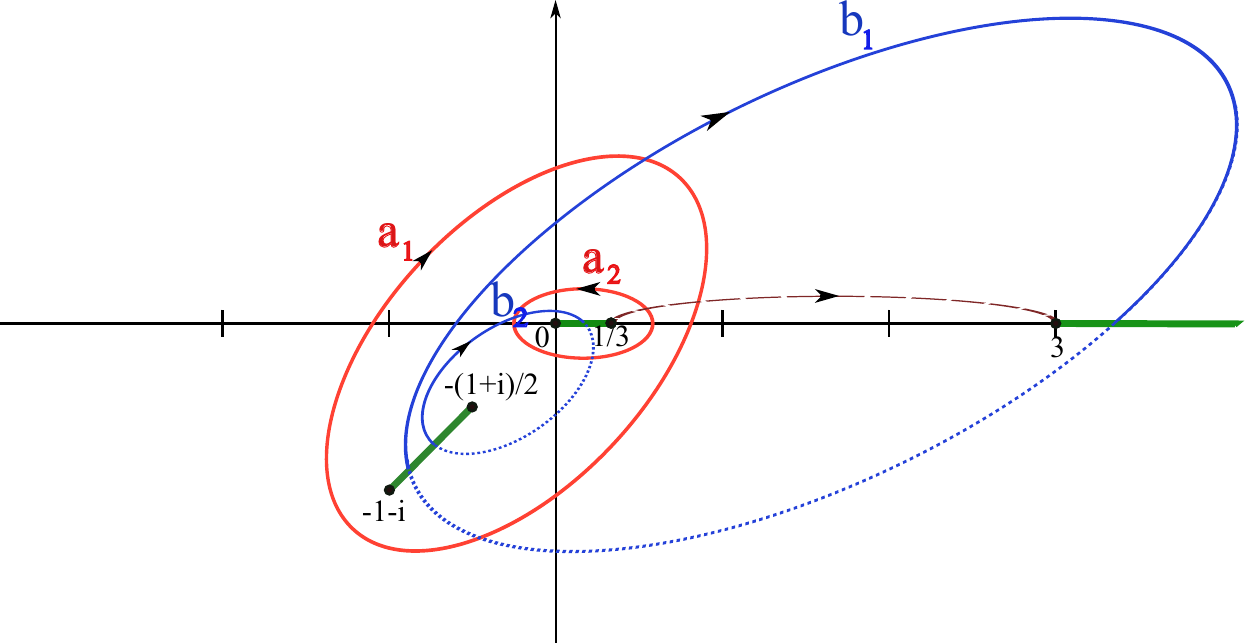}
  \caption{Hyperelliptic Riemann surface for Example 2. Cuts are in green, the brown path is a half period used to define $\hat{\theta}$. }
  \label{cuts2}
  \end{figure}

\begin{figure}
  \centering
  \includegraphics[width=15cm]{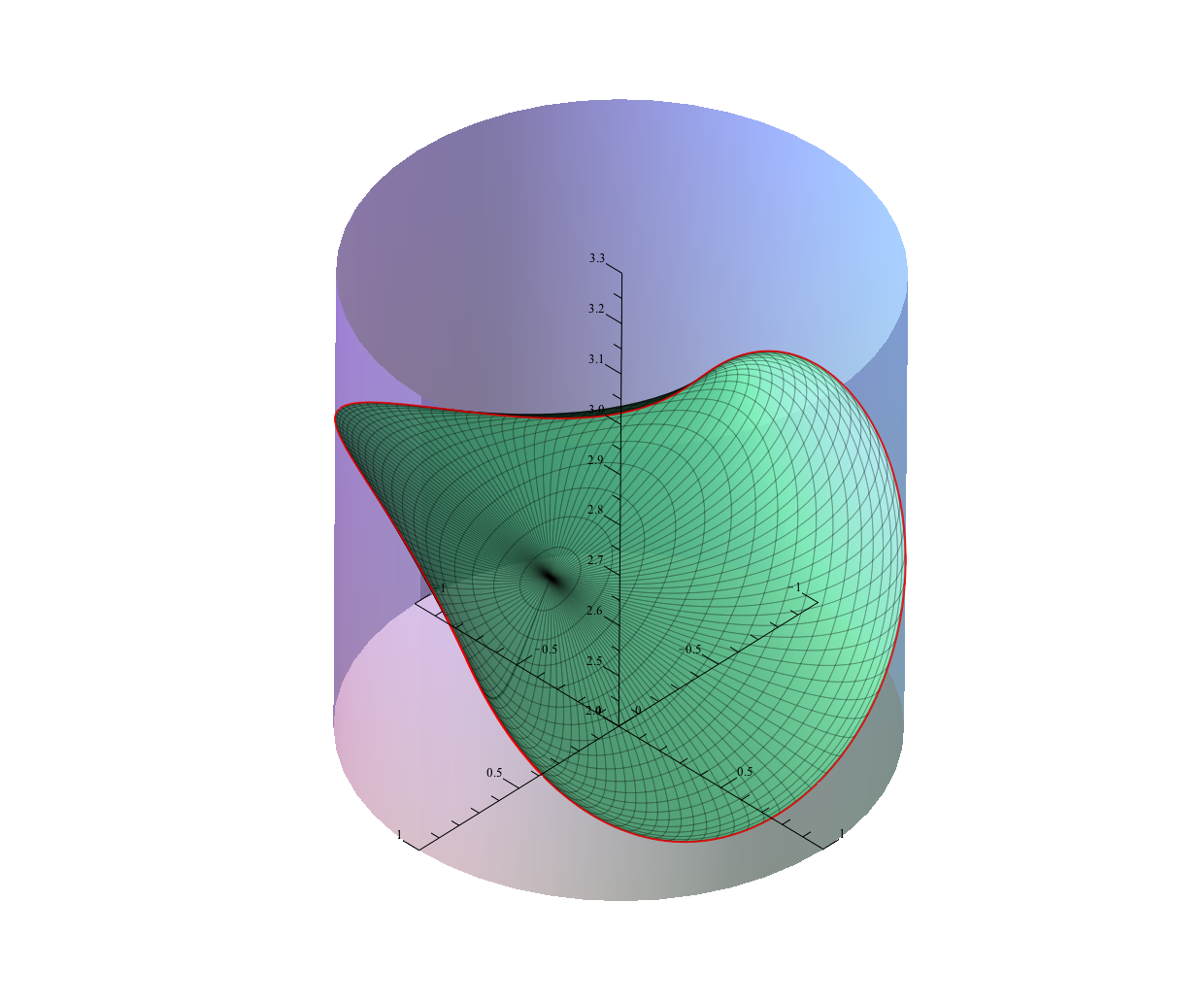}
  \caption{Minimal area surface embedded in \ads{3} in global coordinates $[t,\rho,\phi]$. The vertical direction is time $t$, the radial direction is $\tanh \rho$ and the angle is $\phi$.   }
  \label{surface2}
  \end{figure}

  In this case we choose the branch points at $-1-i, -\half(1+i), 0, \frac{1}{3},3,\infty$ and the basis of cycles is chosen as in figure \ref{cuts2}. The calculations are the same as in the previous example and we just describe the 
values of the relevant quantities as well as depicting the cycles and resulting surface in figs \ref{cuts2} and \ref{surface2}. 
\begin{align}
 \Pi &= \left(\begin{array}{cc} .1837+1.4177 i & .6416 i \\ .6416 i & -.1837+1.4177 i  \end{array}\right) \\
  a &= \half (\mathbb{I} + \Pi) \left(\begin{array}{cc}0 \\ 1\end{array}\right) \\
  p_1 &= \frac{1}{3}, \ \ p_3=3 \\
  \Delta_2 &= \left(\begin{array}{c} 0 \\ 0 \end{array} \right), \ \ \                                      
  \Delta_1 = \left(\begin{array}{c} 1 \\ -1 \end{array} \right) \\
   C_\pm^2 & = \frac{\thetah^2(a)}{D_1\theta(a)D_3\theta(a)} = \frac{4}{5} \sqrt{\frac{2}{17}}\, e^{\frac{3}{4} i\pi} \\ 
 \zeta &= \left(\begin{array}{c} (.1352 + .4419 i) \bar{z}- (0.0284 + 0.0802 i ) z \\ (0.0284 -0.0802 i ) \bar{z}- (.1352 -.4419 i) z \end{array}\right)  \\
 \lambda_4 &= .9987+0.0500 i , \ \ \ \lambda_5 = -.9802-.1982 i \\
 \lv &=  \left(\begin{array}{c} 0.0398 +.1027 i \\ 0.0520-.2854 i \end{array}\right) \\
 \lvt &= \left(\begin{array}{c} -.2098-.1210 i \\ -.6984-.5090 i \end{array}\right) \\ 
 \nu_4 &=   0.3159 - 0.0369 i \\
 \mu_4 &=   1.9562 + 0.3723 i  \\
 \nu_5 &= -0.2647 + 0.0358 i \\
 \mu_5 &=   1.3756 + 0.2997 i \\
 B_1 &= .9939+.1105 i \\
 B_2 &= -.9936-.1132 i \\
\end{align}
In this case the regularized area is given by:
\beq
 A_f = -5.644
\eeq


\section{Acknowledgments}

 We are very grateful to P. Vieira for comments and discussions as well as to J. Toledo for collaboration during the initial stages of this paper. The work of A.I. was supported in part by the Indiana Department of Workforce Development-Carl Perkins Grant and the one of M.K. was supported in part by NSF through a CAREER Award PHY-0952630 
 and by DOE through grant \protect{DE-SC0007884}.

\appendix

\section{Theta function identities}

 In this work we use the notation in $\cite{FK}$, the calculations are similar to those in \cite{KZ,IK}. However, there some small differences, the main one being that $\hat{\theta}(\zeta)$ is defined by an even period and 
 therefore it does not vanish at $\zeta=0$. For that reason we introduced an additional odd half-period $a$ such that $\theta(a)=0$. This modifies the formulas enough that it is
 worth rewriting them. On the other hand the procedure is exactly the same as in \cite{KZ,IK}, namely all identities follow from the quasi-periodicity of the theta function and the fundamental trisecant identity \cite{ThF}, 
 so we do not give detailed derivations. The trisecant identity is
\beq
 \theta(\zeta)\theta(\zeta+\int_j^i+\int_k^l) = \gamma_{ijkl} \theta(\zeta+\int_j^i)\theta(\zeta+\int_k^l) +  \gamma_{ikjl} \theta(\zeta+\int_k^i)\theta(\zeta+\int_j^l)
 \label{trisecant}
\eeq
 with
\beq
 \gamma_{ijkl} = \frac{\theta(a+\int_j^i)\theta(a+\int_l^k)}{\theta(a+\int_l^i)\theta(a+\int_j^k)} 
\label{crossR}
\eeq
 where $a$ is a non-singular zero of the theta function. Now we can take the limit $p_i\rightarrow p_j$ and obtain the first derivative identity
\beq
  D_j\ln\frac{\theta (\zeta)}{\theta (\zeta+\int_k^l)} = D_j\ln\frac{\theta(a-\int_l^j)}{\theta(a+\int_k^j)} -  \frac{D_j\theta(a) \theta(a+\int_l^k)}{\theta (a+\int_l^j)\theta(a+\int_j^k)} \frac{\theta(\zeta+\int_k^j)\theta(\zeta+\int_j^l)}{\theta(\zeta)\theta (\zeta+\int_k^l)} 
\label{1dergen}
\eeq
 Choosing various combination of points $p_{j,k,l}$ the following first derivative identities are obtained
\bal
 D_3\ln\frac{\theta (\zeta)}{\theta (\zeta+\lvo)} &= D_3\ln\frac{\thetah(a-\lvo)}{\thetah(a)} -  \frac{D_3\theta(a)}{\thetah(a)} \frac{\theta (a-\lvo)}{\thetah(a-\lvo)} \frac{\thetah(\zeta)\thetah(\zeta+\lvo)}{\theta(\zeta)\theta (\zeta+\lvo)} \non\\
 D_3\ln\frac{\thetah(\zeta)}{\thetah(\zeta+\lvo)} &= D_3\ln\frac{\thetah(a-\lvo)}{\thetah(a)} - e^{i\pi \Delta_1^t.\Delta_2}  \frac{D_3\theta(a)}{\thetah(a)} \frac{\theta (a-\lvo)}{\thetah(a-\lvo)} \frac{\theta (\zeta)\theta (\zeta+\lvo)}{\thetah(\zeta)\thetah(\zeta+\lvo)} \non\\
 D_1\ln\frac{\theta (\zeta)}{\thetah(\zeta+\lvo)} &= D_1\ln\frac{\theta (a-\lvo)}{\thetah(a)} -  \frac{D_1\theta(a)}{\thetah(a)} \frac{\thetah(a-\lvo)}{\theta (a-\lvo)} \frac{\thetah(\zeta)\theta (\zeta+\lvo)}{\theta (\zeta)\thetah(\zeta+\lvo)} \non\\
 D_1\ln\frac{\thetah(\zeta)}{\theta (\zeta+\lvo)} &= D_1\ln\frac{\theta (a-\lvo)}{\thetah(a)} -  \frac{D_1\theta(a)}{\thetah(a)} \frac{\thetah(a-\lvo)}{\theta (a-\lvo)} \frac{\theta (\zeta)\thetah(\zeta+\lvo)}{\thetah(\zeta)\theta (\zeta+\lvo)} \non \\ \label{firstder}
\end{align}
They can be combined with the trisecant identity (\ref{trisecant}) to obtain, for example
\beq
D_3\ln\frac{\thetah(\zeta+\lvo)}{\thetah(\zeta-\lvo)} = D_3\ln\frac{\thetah(a+\lvo)}{\thetah(a-\lvo)} + e^{i\pi \Delta_1^t.\Delta_2}   \frac{D_3\theta(a)\theta(a-2\lvo)}{\thetah^2(a-\lvo)}  \frac{\theta^2(\zeta)}{\thetah(\zeta+\lvo)\thetah(\zeta-\lvo)} 
\label{D3id}
\eeq
Second derivatives can be obtained similarly, for example, from the first equation in (\ref{firstder}) we obtain, by taking
derivative with respect to $p_4$:
\begin{multline}
D_{43} \ln \theta(\zeta+\lvo) = D_{43}\ln \thetah(a-\lvo) \\ +\frac{D_3\theta(a)\thetah(\zeta)}{\thetah(a)\theta(\zeta)} \frac{\theta(a-\lvo)\thetah(\zeta+\lvo)}{\thetah(a-\lvo)\theta(\zeta+\lvo)} D_4\ln\frac{\thetah(a-\lvo)\thetah(\zeta+\lvo)}{\theta(a-\lvo)\theta(\zeta+\lvo)}
\label{2der}
\end{multline}
Now we can take the limit $p_4\rightarrow p_1$ to obtain
\beq
D_{13}\ln\theta(\zeta) = D_{13} \ln \thetah(a) - \frac{D_3\theta(a)D_1\theta(a)}{\thetah^2(a)}\frac{\thetah^2(\zeta)}{\theta^2(\zeta)}
\label{D13id}
\eeq 
and similarly
\beq
D_{13}\ln\thetah(\zeta) = D_{13} \ln \thetah(a) - e^{i\pi \Delta_1^t.\Delta_2}  \frac{D_3\theta(a)D_1\theta(a)}{\thetah^2(a)}\frac{\theta^2(\zeta)}{\thetah^2(\zeta)}
\label{D13idB}
\eeq 
They can be combined into
\beq
 D_{13} \ln \frac{\theta(\zeta)}{\thetah(\zeta)} =   - \frac{D_3\theta(a)D_1\theta(a)}{\thetah^2(a)}\left[ \frac{\thetah^2(\zeta)}{\theta^2(\zeta)} - e^{i\pi \Delta_1^t.\Delta_2}  \frac{\theta^2(\zeta)}{\thetah^2(\zeta)} \right]
\label{D13idC}
\eeq 
that becomes the sinh-Gordon equation in the main text. The reason is that one takes
\beq
 \zeta= C_{\pm} (\omega_1 z + \omega_3 \bar{z}) 
\label{zetaB}
\eeq
 implying that
\beq
 \partial_z F(\zeta) = C_{\pm} D_1 F(\zeta), \ \ \ \ \bar{\partial}_{\bar{z}} F(\zeta) = C_{\pm} D_3 F(\zeta)
\label{dzC}
\eeq
 where $C_\pm$ is a constant defined in eq.(\ref{e12}). 

 Other useful identity can be obtained from (\ref{1dergen}) by taking $p_j=p_3$, $p_l=p_4$ and expanding for $p_k\rightarrow p_3$. The first non-trivial order gives 
\begin{multline}
\frac{D_3^3\theta(a)}{D_3\theta(a)} - \left(\frac{D_3^2\theta(a)}{D_3\theta(a)}\right)^2  = D_3^2 \ln\left[\thetah(\zeta+\lvo)\theta(\zeta)\thetah(a-\lvo)\right]  \\ + \left(D_3\ln\frac{\theta(\zeta)}{\thetah(\zeta+\lvo)\thetah(a-\lvo)}\right)^2  
    + \frac{D_3^2\theta(a)}{D_3\theta(a)} D_3 \ln\frac{\theta(\zeta)}{\thetah(\zeta+\lvo)\thetah(a-\lvo)}  \label{D^3A}
\end{multline}
Eq. (\ref{D13idC}) together with the identities in eq.(\ref{firstder}) is all that is needed to check the equations of motion. However we are also interested in computing the Schwarzian derivative of the boundary contour. 
This is a more involved calculation for which we derive several identities in the next subsection. 

\subsection{Identities at the world-sheet boundary}

The previous identities are valid for any vector $\zeta\in \mathbb{C}^g$. Since the points at the boundary of the world-sheet are zeros of $\thetah$, in this section we derive identities valid when $\zeta=\zeta_s$ is an arbitrary 
zero of $\thetah$, \ie\ $\thetah(\zeta_s)=0$. From (\ref{firstder}) we immediately get
\begin{align}
 D_3\ln\frac{\theta (\zeta_s)}{\theta (\zeta_s+\lvo)} &= D_3\ln\frac{\thetah(a-\lvo)}{\thetah(a)} \label{bid1} \\
 D_1\ln\frac{\theta (\zeta_s)}{\thetah(\zeta_s+\lvo)} &= D_1\ln\frac{\theta (a-\lvo)}{\thetah(a)} \label{bid2}
\end{align}
from where we find
\begin{align}
 D_1 \ln \frac{\thetah(\zeta_s+\lvo)}{\thetah(\zeta_s-\lvo)} &=  2  D_1 \ln \frac{\thetah(a)}{\theta(a-\lvo)}
 = -\frac{2}{C_\pm} \mu_-  \label{bid3} \\
 D_3 \ln \frac{\theta (\zeta_s+\lvo)}{\theta (\zeta_s-\lvo)} &=  2  D_3 \ln \frac{\thetah(a)}{\thetah(a-\lvo)}
  = -\frac{2}{C_\pm} \mu_+ \label{bid4}
\end{align}
Taking derivative with respect to $p_4$ in identity (\ref{bid1}) we find\footnote{One could also take derivative with respect to $p_1$ but then one has to be careful with a hidden dependence on $p_1$ through the definition of $\thetah$.} 
\begin{align}
 D_{43} \ln\theta(\zeta_s+\lvo) &= D_{43} \ln\thetah(a-\lvo)  \label{bid6}\\
 D_{13} \ln\theta(\zeta_s) &= D_{13} \ln\thetah(a)  \label{bid5}
\end{align}
where, in the second one we also took the limit $p_4\rightarrow p_1$. 
Multiplying the second and fourth equations in (\ref{firstder}) by $\thetah(\zeta)$ and taking $\zeta=\zeta_s$ it follows that
\begin{align}
 D_3 \thetah(\zeta_s) &= - e^{i\pi \Delta_1^t.\Delta_2} \frac{D_3\theta(a)\theta(a-\lvo)}{\thetah(a)\thetah(a-\lvo)} \frac{\theta(\zeta_s)\theta(\zeta_s+\lvo)}{\thetah(\zeta_s+\lvo)}  \label{bid8} \\
 D_1 \thetah(\zeta_s) &= - \frac{D_1\theta(a)\thetah(a-\lvo)}{\thetah(a)\theta(a-\lvo)} \frac{\theta(\zeta_s)\thetah(\zeta_s+\lvo)}{\theta(\zeta_s+\lvo)}  \label{bid8b} 
\end{align}
Also, multiplying the second equation in (\ref{firstder}) by $\thetah(\zeta)$ taking derivative $D_3$ with respect to $\zeta$ and setting $\zeta=\zeta_s$ it follows that
\beq
\frac{D_3^2\thetah(\zeta_s)}{D_3\thetah(\zeta_s)} = 2D_3\ln\theta(\zeta_s)  \label{bid9}
\eeq
where (\ref{bid1}) was used to simplify the result. Taking derivative $D_3$ with respect to $\zeta$ in the third equation in (\ref{firstder}), taking $\zeta=\zeta_s$, and using (\ref{bid8b}) we obtain
\beq
 D_{13}\ln \frac{\theta(\zeta_s)}{\thetah(\zeta_s+\lvo)} = \left(\frac{D_1\theta(a)\thetah(a-\lvo)}{\thetah(a)\theta(a-\lvo)}\right)^2\ \frac{D_3\thetah(\zeta_s)}{D_1\thetah(\zeta_s)} \label{bid7}
\eeq
Taking derivative $D_3$ with respect to $\zeta$ in the first equation in (\ref{firstder}), taking $\zeta=\zeta_s$, and using (\ref{bid8}) we obtain
\beq
 D_3^2 \ln \frac{\theta(\zeta_s)}{\theta(\zeta_s+\lvo)} = e^{i\pi \Delta_1^t.\Delta_2} \frac{\lambda}{C_\pm^2} 
  \label{bid10}
\eeq
where we replaced
\beq
 \lambda = C_\pm^2\left(\frac{D_3\theta(a) \theta(a-\lvo)}{\thetah(a)\thetah(a-\lvo)}\right)^2
\label{leqB}
\eeq
as follows from the definitions of $C_\pm$ and $\lambda$, \ie\ eqs.(\ref{e12}) and (\ref{e19}).

Finally, multiplying the second equation in (\ref{firstder}) by $\thetah(\zeta)$ taking second derivative $D^2_3$ with respect to $\zeta$ and setting $\zeta=\zeta_s$ it follows that
\begin{multline}  
 D_3^2\ln\frac{\thetah(\zeta_s+\lvo)}{\theta(\zeta_s)} + \left(D_3\ln\frac{\thetah(a-\lvo)\thetah(\zeta_s+\lvo)}{\thetah(a)\theta(\zeta_s)}\right)^2 =  \\ \frac{D_3^3\thetah(\zeta_s)}{D_3\thetah(\zeta_s)} - 3D_3^2\ln\theta(\zeta_s) 
    -3 \left(D_3\ln\theta(\zeta_s)\right)^2 + 
 e^{i\pi \Delta_1^t.\Delta_2}\frac{\lambda}{C_\pm^2} \label{bid11} 
\end{multline}
where eqs.(\ref{bid9}), (\ref{bid1}), (\ref{bid10}) were used to simplify the result. An equation similar to the last one can be derived by simply setting $\zeta=\zeta_s$ in eq.(\ref{D^3A}). The results agree only if 
\begin{multline}
\frac{D_3^3\thetah(\zeta_s)}{D_3\thetah(\zeta_s)} - \frac{D_3^2\theta(\zeta_s)}{\theta(\zeta_s)} - 2\left(\frac{D_3\theta(\zeta_s)}{\theta(\zeta_s)}\right)^2 = \\ 
\frac{D_3^3\theta(a)}{D_3\theta(a)} - \left(\frac{D_3^2\theta(a)}{D_3\theta(a)}\right)^2+\frac{D_3^2\theta(a)D_3\thetah(a)}{D_3\theta(a)\thetah(a)} - \frac{D_3^2\thetah(a)}{\thetah(a)}
\label{bid12}
\end{multline}
which is the last identity we need. It is equivalent to say that the left hand side is independent of the zero if $\thetah$ that we take. In particular if we take $\zeta_s = a +\int_1^3$ we obtain the right-hand side. 

\subsection{Identities at particular points}

 One last type of identity is needed in order to fix the spectral parameter $\lambda$ to any desired value. Indeed, according to eq.(\ref{e19}), $\lambda$ is obtained by first choosing a point $p_4$ and then computing 
\beq
 \lambda(p_4) = \frac{D_3\theta(a)}{D_1\theta(a)} \left( \frac{\theta(a-\int_1^4)}{\thetah(a-\int_1^4)} \right)^2
 \label{pid1}
\eeq
In practice we fix first $\lambda$ and then choose $p_4$ accordingly, namely we need to invert the function $\lambda(p_4)$.  The main observation is that the right hand side of the equation, as a function of $p_4$, has the following properties: it is a well-defined function on the Riemann surface, namely independent of the path used to define the integral $\int_1^4$. Second it takes the same values on both sheets of the Riemann surface, namely it has no cuts and therefore it is a well-defined function of $\lambda_4$, the projection of $p_4$ onto the complex plane. Finally, as a function of $\lambda_4$ it has a zero at $\lambda=\lambda_1$ and a pole at $\lambda_4=\lambda_3$ (where $\lambda_{1,3}$ are the branch points taken to be $p_{1,3}$). It has no other zeros or poles. This last property is perhaps the only that requires an explanation since, as function of $p_4$ the theta functions in the numerator and denominator have $g-1$ additional zeros. The fact is that all those zeros coincide and therefore cancel between numerator and denominator. 
This can be checked \cite{FK,ThF} using Riemann's theorem to write $a=\kappa+\int_{p_1}^{q_1}+\ldots+\int_{p_1}^{q_{g-1}}$ where $\kappa$ is the Riemann constant and
$q_{1\ldots g-1}$ are $g-1$ points on the Riemann surface that turn out also to be the zeros of the numerator and denominator.  
 Taking into account all these properties, we can write
\beq
 \lambda = \frac{D_3\theta(a)}{D_1\theta(a)} \left( \frac{\theta(a-\int_1^4)}{\thetah(a-\int_1^4)} \right)^2 = A_0 \frac{\lambda_4-\lambda_1}{\lambda_4-\lambda_3}
\label{pid2}
\eeq
for some constant $A_0$. This constant can be evaluated by considering the limits $\lambda_4\rightarrow \lambda_1$ and $\lambda_4\rightarrow \lambda_3$. We obtain
\beq
 A_0 = -\frac{1}{C_\pm^2} \frac{4 \lambda_1^{2g-2}}{\prod_{i\neq 1,3}(\lambda_1-\lambda_i)} 
     = -e^{i\Delta_1^t\Delta_2}\, \frac{C_\pm^2}{4} \lambda_3 \prod_{i\neq 1,3}(\lambda_3-\lambda_i)
\label{pid3}
\eeq
where the products are over all branch points except $p_1$, $p_3$, $0$ and $\infty$. Since the two expressions for $A_0$ have to agree we find that
\beq
  C_\pm^4 =  \frac{16}{\lambda_1\lambda_3}\ \frac{ \lambda_1^{2g-2} e^{-i\pi\Delta_1^t\Delta_2}}{\prod_{i\neq1,3}[(\lambda_1-\lambda_i)(\lambda_3-\lambda_i)]}
\label{pid4}
\eeq
 Finally we get, for the spectral parameter $\lambda$
\beq
 \lambda = \pm i |\lambda_3| e^{i\pi\Delta_1^t\Delta_2}
 \left(\frac{\lambda_1}{\bar{\lambda}_1}\right)^{g-1} \prod_{i\neq1,3}\left(\frac{\sqrt{\lambda_i}\,\sqrt{\bar{\lambda}_1-\bar{\lambda}_i}}{\sqrt{\lambda_1-\lambda_i}}\right) \ \frac{\lambda_4-\lambda_1}{\lambda_4-\lambda_3}
\label{pid5}
\eeq
 which allows us to easily choose $\lambda_4$ to obtain any $\lambda$ we desire.  In fact it is easily seen that $|\lambda|=1$ if and only if $|\lambda_4|=1$, thus for real solutions we just take $\lambda_4$ on the unit circle.

\subsection{Schwarzian derivative}

 The formulas summarized in the previous subsections can be used to derive a particularly simple expression 
for the Schwarzian derivative of the contour similar to the one found in \cite{KZ}. We will be using that, from eq.(\ref{dzC})
it follows that
\begin{align}
 \partial_s F(\zeta_s) &= \partial_s z\ \partial_z F(\zeta_s) +  \partial_s \bar{z}\ \partial_{\bar{z}} F(\zeta_s) \label{Sd0a} \\
 & = C_\pm  \partial_s z\ D_1 F(\zeta_s) + C_\pm \partial_s \bar{z}\ D_3 F(\zeta_s)
\label{Sd0}
\end{align}
In particular since $\thetah(\zeta_s)=0$, we obtain
\beq
  \partial_s z\ D_1 \thetah(\zeta_s) + \partial_s \bar{z}\ D_3 \thetah(\zeta_s) = 0
\label{Sd0B}
\eeq
which determines the direction tangent to the world-sheet contour $(z(s),\bar{z}(s))$. 
Now, starting from 
\beq
 \hat{x}_+ = - e^{-2\mu_4 z-2\nu_4\bz}\, \frac{\hat{\theta}(\zeta_s-\lv)}{\hat{\theta}(\zeta_s+\lv)}
 \label{Sd1}
\eeq
 we obtain using eqs.(\ref{Sd0}) and (\ref{bid3}):
\beq
 \partial_s \ln \hat{x}_+ = -\partial_s \bar{z} \left(2\mu_+ + C_\pm D_3\ln\frac{\thetah(\zeta_s+\lvo)}{\thetah(\zeta_s-\lvo)}\right)
\label{Sd2}
\eeq
Then, thanks to eq.(\ref{bid1}) we find
\beq
 \frac{\partial_s^2 \hat{x}_+}{\partial_s \hat{x}_+} = \frac{\partial_s^2 \bar{z}}{\partial_s \bar{z}} -2 \mu_+ \partial_s \bar{z} + 2 C_{\pm} \partial_s \bar{z} D_3 \ln\frac{\theta(\zeta_s)}{\thetah(\zeta_s+\lvo)}
\label{Sd3}
\eeq
Now it is straight-forward to compute $\{\hat{x}_+,s\}=\{x_+,s\}$ and then simplify the result using eqs.(\ref{bid7}) and (\ref{bid11}) to obtain:
\beq
\{x_+,s\} = \{\bar{z},s\} - 2 \lambda_1 (\partial_s\bar{z})^2 - \frac{2}{\lambda_1} (\partial_s \bar{z})^2 
+ 2 C_\pm^2  (\partial_s\bar{z})^2  \left[-\frac{D_3^3\thetah(\zeta_s)}{D_3\thetah(\zeta_s)}+3\frac{D_3^2\theta(\zeta_s)}{\theta(\zeta_s)}\right]
\label{Sd4}
\eeq
 Further simplification using eq.(\ref{bid12}) results in: 
\beq
 \{x_+,s\} = \{\bar{z},s\} - 2 \lambda (\partial_s\bar{z})^2 - \frac{2}{\lambda} (\partial_s \bar{z})^2 + 2 C_\pm^2 
 (\partial_s\bar{z})^2 \left[ 2 D_3^2 \ln\theta(\zeta_s) - C_{sd} \right]
\label{Sd12}
\eeq
with $C_{sd}$ a constant given by
\beq
 C_{sd} = \frac{D_3^3\theta(a)}{D_3\theta(a)} - \left(\frac{D_3^2\theta(a)}{D_3\theta(a)}\right)^2+\frac{D_3^2\theta(a)D_3\thetah(a)}{D_3\theta(a)\thetah(a)} - \frac{D_3^2\thetah(a)}{\thetah(a)}
\label{Sd13}
\eeq
as used in the main text. Notice also that $x_-$ has the same expression as $x_+$ except that the point $p_4$ in the Riemann surface is replaced by $p_5$. In the Schwarzian derivative the only effect is to replace $\lambda\rightarrow -\lambda$. Thus
\beq
 \{x_-,s\} = \{\bar{z},s\} + 2 \lambda (\partial_s\bar{z})^2 + \frac{2}{\lambda} (\partial_s \bar{z})^2 + 2 C_\pm^2 
 (\partial_s\bar{z})^2 \left[ 2 D_3^2 \ln\theta(\zeta_s) - C_{sd} \right]
\label{Sd14}
\eeq

\end{document}